\documentclass[a4paper,fleqn,usenatbib]{mnras}

\usepackage{newtxtext,newtxmath}

\usepackage[T1]{fontenc}
\usepackage{ae,aecompl}

\usepackage{graphicx}	
\usepackage{amsmath}	
\usepackage{amssymb}	

\newcommand\U[1]{{\,\rm #1}}
\newcommand\E[1]{\times10^{#1}}

\DeclareMathOperator{\erfc}{erfc}
\DeclareMathOperator{\erfi}{erfi}
\newcommand\al{\alpha}
\newcommand\bt{\beta}
\newcommand\gm{\gamma}
\newcommand\dl{\delta}
\newcommand\tht{\theta}
\newcommand\eps{\epsilon}
\newcommand\kp{\kappa}
\newcommand\lmb{\lambda}
\newcommand\sg{\sigma}
\newcommand\Dl{\Delta}
\newcommand\Sg{\Sigma}
\newcommand\rs[1]{_\mathrm{#1}}
\newcommand\Vsh{V\rs{sh}}
\newcommand\Rcurv{R\rs{curv}}
\newcommand\Zsh{Z\rs{sh}}
\newcommand\FDawson{F\rs{D}}
\newcommand\hexp{h\rs{exp}}
\newcommand\dtwo{d_2}
\newcommand\DlV{\Dl V}
\newcommand\Lbt{{\mathrm{Ly}}\!\bt}
\newcommand\Hal{{\mathrm H}\al}
\newcommand\Hbt{{\mathrm H}\bt}
\newcommand\Hgm{{\mathrm H}\gm}
\newcommand\kms{km\,s^{-1}}
\newcommand\zmax{z\rs{max}}
\newcommand\lmbLOS{\lmb\rs{LOS}}

\title[Balmer filament in SN 1006]{Interplay between Physics and Geometry in Balmer filaments: the Case of SN 1006}

\author[R. Bandiera, et al.]{
R. Bandiera$^{1}$,\thanks{E-mail: bandiera@arcetri.astro.it (RB)},
G. Morlino$^{2,1,3}$,
S. Kne\v{z}evi\'c$^{4}$,
J.C. Raymond$^{5}$\\
$^{1}$INAF - Osservatorio Astrofisico di Arcetri, Largo E. Fermi 5,
I-50125 Firenze, Italy\\
$^{2}$Gran Sasso Science Institute, Viale F.~Crispi 7, L'Aquila, Italy \\
$^{3}$INFN/Laboratori Nazionali del Gran Sasso, Via G.~Acitelli 22, Assergi (AQ), Italy \\
$^{4}$Astronomical Observatory Volgina 7, 11060 Belgrade, Republic of Serbia\\
$^{5}$Harvard-Smithsonian Center for Astrophysics, 60 Garden St., Cambridge, MA 02138, USA
}

\date{Accepted 2018 November 21. Received 2018 November 13; in original form 2018 July 5}
\pubyear{2018}

\begin{document}
\label{firstpage}
\pagerange{\pageref{firstpage}--\pageref{lastpage}}
\maketitle

\begin{abstract}
The analysis of Balmer-dominated emission in supernova remnants is potentially a very powerful way to derive information on the shock structure, on the physical conditions of the ambient medium and on the cosmic-ray acceleration efficiency.
However, the outcome of models developed in plane-parallel geometry is usually not easily comparable with the data, since they often come from regions with rather a complex geometry.
We present here a general scheme to disentangle physical and geometrical effects in the data interpretation, which is especially powerful when the transition zone of the shock is spatially resolved and the spectral resolution is high enough to allow a detailed investigation of spatial changes of the line profile.
We then apply this technique to re-analyze very high quality data of a region along the northwestern limb of the remnant of SN~1006.
We show how some observed features, previously interpreted only in terms of spatial variations of physical quantities, naturally arise from geometrical effects. With these effects under control, we derive new constraints on physical quantities in the analyzed region, like the ambient density (in the range 0.03--$0.1\U{cm^{-3}}$), the upstream neutral fraction (more likely in the range 0.01--0.1), the level of face-on surface brightness variations (with factors up to $\sim 3$) and the typical scale lengths related to such variations ($\sog 0.1\U{pc}$, corresponding to angular scales $\sog 10\U{arcsec}$).
\end{abstract}

\begin{keywords}
ISM: supernova remnants -- ISM: individual objects: SN 1006 -- shock waves -- radiation mechanisms: thermal
\end{keywords}



\section{Introduction}	\label{sec:intro}

There is a rather general consensus on the fact that Supernova Remnants (SNRs) are the most likely candidates for the acceleration of Galactic Cosmic Rays, up to energies of the order of $10^{15}\U{eV}$.
However, the details of how this acceleration proceeds, for ions as well as for electrons, have not been assessed yet with sufficient confidence: for instance, efficient diffusive acceleration requires a strong turbulent amplification of the magnetic field, and consequently a dynamical feedback of cosmic rays (CRs) and magnetic fields on the shock structure itself.
So, one of the ways to test the presence of efficient shock acceleration is to find clues of a CR-modified shock like, for instance, the evidence of a precursor, or of a thermal energy sink in the downstream, or even of concavities in the electron synchrotron spectrum.

In this sense, Balmer-dominated shock emission offers a very important diagnostic tool.
It appears when a non-radiative shock moves through a partially neutral medium \citep{ChevalierRaymond1978, Chevalier+1980}.
Its theoretical grounds are rather well assessed, because they rely on only a few collisional processes; excitation, ionization and charge exchange. In the most basic picture, some of the neutral hydrogen atoms entering the shock are collisionally excited, and emit narrow Balmer lines, whose width is related to the kinetic temperature of the upstream neutrals; part of the original neutrals undergo instead a charge-exchange process, and then the new fast moving neutrals (formerly being shocked ions) will emit broad Balmer lines; these second-generation neutrals have some chance to undergo charge exchange again, and so to create further generations of neutrals, until all neutrals will eventually get ionized.
Due to this chain of processes, Balmer lines are expected to show a broad and a narrow component \citep{Chevalier+1980}, with their relative strengths depending on the relative importance of collisional ionization and charge-exchange reaction rates.

In spite of the simple physical processes involved, a detailed model implementation is rather difficult: firstly because we need to follow a complex reaction tree \citep[see e.g.][]{HengMcCray+2007}; but also because the mean free path of neutral atoms is larger or comparable to the shock thickness hence they never behave like a fluid.
Therefore their local velocity distributions may strongly deviate from a Maxwellian and, in addition, some of them may overtake the shock front and reach the upstream medium, then giving rise to a precursor even in the absence of CRs \citep{paperI, Hester+1994, Lim+1996}.

Neutral atoms are not directly involved in the CR acceleration process.
Nonetheless, the properties of Balmer lines could be affected, even though in a complex way, by the presence of efficient CR acceleration, and therefore Balmer emission could represent an effective diagnostic tool to search for the presence of CRs. In this respect there are three main observables that can provide hints on the CR presence:
1) detection of $\Hal$ emission from the region immediately ahead of the shock that could result from the presence of a CR precursor able to heat the upstream plasma \citep[e.g.][]{Lee+2007, Lee+2010, Katsuda+2016}; 
2) broadening of the narrow line component \citep[first seen by][]{Smith+1994} also due to the formation of a CR precursor;
3) a reduction of the broad line width resulting from the energy transfer from the downstream thermal bath to the CR component (see e.g., for SNR~0509--67.5, \citealt{Helder+2010} but also see \citealt{paperIV}; and, for RCW~86, \citealt{hea13} but also \citealt{RCW86}).

Such a scenario is further complicated by the fact that Balmer emission also depends on the temperature equilibration level between electrons and ions \cite[e.g.][]{Smith+1991,paperII}, a quantity hard to derive {\it a priori} from theory, but on which observations suggest a very clear trend \citep{Ghavamian+2007, Ghavamian+2013}.

In this paper we will tackle the problem under a different perspective.
More recent and detailed observations are able to resolve, or partially resolve, the spatial structure of the shock layer from which Balmer emission is emitted, and a comparison of these data with models may provide much more information than low resolution observations.
Nevertheless, in order to effectively link theory and observations, it is not sufficient to develop very sophisticated models and to derive 1-D profiles, just as in the case of a pure plane parallel shock seen edge-on.
Instead, actual observations may refer to more complex geometries like curved or even rippled shocks which result in multiple layers seen on a single line of sight (LOS).
Therefore, in general, one must be aware of the importance of a correct understanding of the geometry, before attempting a physical interpretation of the observations of Balmer filaments.

The present work has two main motivations.
First of all, we aimed at outlining a general scheme for linking models and observations.
To this purpose we introduced two levels of simplified models, the former of which closely mimics the results obtained with the kinetic code developed in our previous works, while the latter one relies on parameters that can be more directly derived from observations.
Then, we analyzed the relation between 1-D spatial models and the projected spatial profiles of actual observations, for different cases of curved shock layers.

Our second aim is to apply this scheme to a well known and very deeply observed region of a filament lying along the northwestern side of the remnant of SN~1006, in order to test its diagnostic power, as well as to obtain novel determinations of some physical parameters in that region.

The plan of the paper is the following: in Section 2 we review our previous work on kinetic models of shocks in partially neutral media and related Balmer emission; in Section 3 we introduce two levels of analytic models, labelled as ``3-fluid model'' and ``parametric model'' respectively, which will serve at a bridge between the kinetic models and actual observations; Section 4 presents a general method to relate plane-parallel profiles to the projected profiles, when the shock is seen edge-on and presents a curvature; then in Section 5, by using our parametric model, we discuss in more detail the expected properties of spatially resolved profiles, for different cases of shock curvature; in Section 6 we review some recent and very detailed observations of Balmer emission along the northwestern limb of the remnant of SN~1006 \citep{Raymond+2007, Nikolic+2013}; Section 7 is devoted to a re-analysis of the data by \citet{Raymond+2007}, in which we exploit the superb spatial resolution of HST data; in Section 8, instead, we analyze the \citet{Nikolic+2013} $\Hal$ data, with lower spatial resolution but with complete characterization of the line profiles. We not only test their consistency with the HST data discussed in the previous section, but we also discuss the diagnostic potential of measurements of width and offset of the broad-line component in several locations;  in Section 9 we compare our density estimates with those present in the literature, and discuss strengths and weaknesses of the various approaches; Section 10 concludes.

\section{Shocks in partially neutral media}  \label{sec:model}

Here we summarize the kinetic model for shock particle acceleration in the presence of neutrals developed in \cite{paperI, paperII, paperIII}. 
We consider a stationary system with a plane-parallel shock wave propagating in a partially ionized proton-electron plasma with velocity $\Vsh$ along the $z$ direction. The fraction of neutral hydrogen is fixed at upstream infinity where ions and neutrals are assumed to be in thermal equilibrium with each other. The shock structure is determined by the interaction of CRs and neutrals with the background plasma. Both CRs and neutrals may profoundly change the shock structure, especially upstream where both create a precursor: the CR-induced precursor reflects the diffusion properties of accelerated particles and has a typical spatial scale of the order of the diffusion length of the highest energy particles. The neutral-induced precursor develops on a spatial scale comparable with a few interaction lengths of the dominant process between CE and ionization. The downstream region is also affected by the presence of both CRs and neutrals and the velocity gradients that arise from ionization have a direct influence on the spectrum of accelerated particles.
A self consistent description of shock particle acceleration in the presence of neutral hydrogen requires the consideration of four mutually interacting species: thermal particles (protons and electrons), neutrals (hydrogen), accelerated protons (CRs) and turbulent magnetic field. We neglect the presence of helium and heavier chemical elements. This is a good approximation because there is little exchange of  energy among different ion species in fast shocks \citep{Korreck+2004, Raymond+2017}.

Let us start with the description of neutrals. The main difficulty arises from the fact that neutrals cannot be described as a fluid, because in the downstream the collisional ionization length is smaller than the equilibration length. Hence neutrals are described kinetically, using  the stationary Boltzmann equation to calculate the evolution of the velocity distribution function, $f_N(\vec v,z)$,
\begin{equation}
\label{eq:vlasov}
v_z \frac{\partial f_{N}(\vec v, z)}{\partial z} = \beta_{N} f_{p}(\vec v, z)  -
        \left[ \beta_{p} + \beta_{e} \right] f_{N}(\vec v, z) \,,
\end{equation}
where $z$ is the distance from the shock (which is located at the origin), taken as positive in the downstream, $v_z$ is the velocity component along the $z$ axis and the electron and proton distribution functions, $f_{e}(\vec v,z)$ and $f_{p}(\vec v,z)$, are assumed to be Maxwellian at each position. The collisional terms in Eq.~(\ref{eq:vlasov}), $\beta_k f_l$, describe the interaction (due to CE and/or ionization) between the species $k$ ($=i$, $e$, $N$) and $l$ (= $i$, $N$). The interaction rate $\beta_k$ is formally written as
\begin{equation} \label{eq:beta_k}
\beta_k (\vec v,z) = \int d^{3} w \, v\rs{rel} \, \sg(\vec v\rs{rel})
                  f_{k}(\vec w,z) \,,
\end{equation}
where $v\rs{rel} = |\vec v- \vec w|$ and $\sg$ is the cross section for the
relevant interaction process. More precisely, $\beta_N$ is the rate of CE of an
ion that becomes a neutral, $\beta_{p}$ is the rate of CE plus ionization of a
neutral due to collisions with protons, while $\beta_{e}$ is the ionization rate
of neutrals due to collisions with electrons. A full description of the cross
sections used in the calculations can be found in \cite{paperII}.

The isotropic distribution function of CRs satisfies the following transport equation in
the reference frame of the shock \citep{Skilling1971,Drury1983}:
\begin{equation} \label{eq:trasp_CR}
 \frac{\partial}{\partial z} \left[ D(z,p) \frac{\partial f}{\partial z}
 \right]
 - u \frac{\partial f}{\partial z} 
 + \frac{1}{3} \frac{d u}{d z} \, p \frac{\partial f}{\partial p} 
 + Q(z,p) = 0  \,,
\end{equation}
where $D(z,p)$ is the diffusion coefficient and $Q(z,p)$ ia the injection term.
The $z$-axis is oriented from upstream infinity $(z=-\infty)$ to downstream
infinity $(z=+\infty)$ with the shock located at $z=0$. We assume that the
injection occurs only at the shock position and is monoenergetic at $p=p_{\rm
inj}$.
The diffusion properties of particles are described by $D(z,p)$. We assume Bohm diffusion in the local amplified magnetic
field:
\begin{equation} \label{eq:Diff}
 D(z,p) = \frac{1}{3} c r_L[\delta B(z)] \, ,
\end{equation}
where $r_L(\delta B)= pc/[e \delta B(z)]$ is the Larmor radius in the amplified magnetic field. The calculation of $\delta B$ is described assuming that the only turbulence which scatters particles is the one self-generated by the particles themselves through the resonant streaming instability. These waves are also damped due to several processes. In particular, when the plasma is not fully ionized, the
presence of neutrals can damp Alfv\`en waves via ion-neutral damping.  
The equation for transport of waves can be written as:
\begin{equation} \label{eq:wave_tr}
  \partial_z F_w=  u(z) \,\partial_z P_w +
   P_w \left[ \sg_{\rm CR}(k,z)- \Gamma_{\rm TH}(k,z)   \right] \,,
\end{equation}
where $F_w(k,z)$ and $P_w(k,z)$ are, respectively, the energy flux and the pressure per unit logarithmic bandwidth of waves with wavenumber $k$. $\sg$ is the growth rate of magnetic turbulence, while $\Gamma_{\rm TH}$ is the damping rate. For resonant wave amplification the growth rate of Alfv\'en waves is:
\begin{equation} \label{eq:sigma_CR}
 \sg_{\rm CR}(k,x)= \frac{4\pi}{3} \frac{v_A(x)}{P_w(k,x)} \left[ 
    p^4 v(p) \frac{\partial f}{\partial x} \right]_{p=\bar p(k)} \,,
\end{equation}
where $p=\bar p(k)= eB/k m_{p} c$ is the resonant momentum. The damping of the waves is mainly due to non-linear Landau damping and ion-neutral damping. For the sake of simplicity here we adopt a phenomenological approach in which the damping results in a generic turbulent heating (TH) at a rate  $\Gamma_{\rm TH} = \eta_{\rm TH} \sg_{\rm CR}$. This expression assumes that a fraction $\eta_{\rm TH}$ of the power in amplified waves is locally damped and results in heating of the background plasma. 

Finally we need to describe the dynamics of the background plasma which is affected by the presence of accelerated particles and by CE and ionization of neutrals. Protons and electrons in the plasma are assumed to share the same local number density, $\rho_{p}(z)/m_{p}=\rho_{e}(z)/m_{e}$, but not necessarily the same temperature, i.e., $T_{p}(z)$ may be different from $T_{e}(z)$. The equations describing the conservation of mass, momentum and energy taking into account the interactions of the plasma fluid with CRs are:

\begin{equation} \label{eq:rh1}
 \frac{\partial}{\partial z} \left[\rho_{p} u_{p} + \mu_N  \right]=0 \,,
\end{equation}
\begin{equation} \label{eq:rh2} 
 \frac{\partial}{\partial z} \left[ \rho_{p} u_{p}^{2} + P_{g} + P_{c} + P_{w} 
        + P_{N}  \right]=0 \,,
\end{equation}
\begin{equation} \label{eq:rh3}
 \frac{\partial}{\partial z} \left[ \frac{1}{2} \rho_{p} u_{p}^{3} + 
  \frac{\gamma_{g} P_{g} u_{p}}{\gamma_{g}-1} + F_w + F_{N} \right]
  = -u_{p} \frac{\partial P_c}{\partial z} + \Gamma P_w \,.
\end{equation}
Here $\mu_N = m_H \int d^{3} v v_{\parallel} f_{N}$, $P_N = m_H \int d^{3} v v_{\parallel}^{2} f_{N}$ and $F_N = m_H/2 \int d^{3} v v_{\parallel} (v_{\parallel}^{2} + v_{\perp}^{2}) f_{N}$ are respectively the fluxes of mass, momentum and energy of neutrals along the $z$ direction (labelled as $\parallel$).  They can be easily computed once the neutral distribution function is known. $P_w$ and $F_w$ are the pressure and energy flux of waves, while $P_c$ is the CR pressure computed from the CR distribution function:
\begin{equation}
 P_c(z) = \frac{4 \pi}{3} \int dp \, p^3 v(p) f(z,p) \,.
\end{equation}
The dynamical role of electrons in the conservation equations is usually neglected due to their small mass. However, collective plasma processes could contribute to equilibrate electron and proton temperatures, at least partially. If the equilibration occurs in a very efficient manner, the electron pressure cannot be neglected and the total gas pressure needs to include both the proton and electron contributions, namely $P_{g} = P_{p} + P_{e} = P_{p}(1+T_e/T_p)$, where $T_e/T_p$ is the electron to proton temperature ratio. While it is well established that electron-proton equilibration in the downstream is partial for Balmer-dominated shocks with velocities exceeding $500\U{km\,s^{-1}}$ \citep{Ghavamian+2001, Rakowski+2003, Ghavamian+2007},  in the presence of a precursor (either induced by the CRs or by the neutrals),  also upstream of the shock the level of equilibration becomes an unknown. Nevertheless, motivated by the fact that in the northwestern filament of SN 1006 there are no strong indications for the presence of a precursor, we fix the upstream electron temperature to $10^4$ K, while the value of $T_e/T_p$ downstream is taken as a free parameter and constant in the whole volume of shocked plasma.

In order to solve the set of non-linear equations involving neutrals, ions, CRs and magnetic field, we adopt an iterative method that is fully described in \cite{paperIII}. The input quantities are the values of the shock velocity and all environmental quantities at upstream infinity, where the distribution function of neutrals is assumed to be Maxwellian at the same temperature as that of ions. 
At the end, the procedure provides the distribution functions of neutral hydrogen, protons and electrons. The subsequent calculation of $\Hal$ emission is described in \cite{paperII}. We first calculate the production rate of hydrogen excited to level $3l$ taking into account both the contributions due to excitation and to charge-exchange as follows
\begin{eqnarray} \label{eq:H(3l)}
 R_{H(3l)}(\vec v,z) = \int d^3 w \, v_{\rm rel} \, f_{N}(\vec v,z) \times \nonumber \\
      \left[ f_p(\vec w,z)  \, \sg_{ex(i)}^{(3l)}(v_{\rm rel}) 
      + f_e(\vec w,z)  \, \sg_{ex(e)}^{(3l)}(v_{\rm rel}) \right]  \nonumber \\
   + \int d^3 w \, v_{\rm rel} \,
      f_p(\vec v,z) f_{N}(\vec w,z) \, \sg_{ce}^{(3l)}(v_{\rm rel}) \,.
\end{eqnarray}
Finally, the total production rate of $\Hal$ photons as a function of the position $z$ and velocity $\vec {v}$ is given by
\begin{equation} \label{eq:Ralpha}
 R_{\Hal} = R_{{\mathrm H}(3s)} + R_{{\mathrm H}(3d)} + B_{3p,2s} R_{{\mathrm H}(3p)} \,.
\end{equation}
The factor $B_{3p,2s}$ is the fraction of transitions from $3p$ to $2s$, which is $\approx 0.12$ in the optically thin case (Case A) while it becomes unity in the optically thick case (Case B) \citep{vanAdelsberg+2008}.
\citet{Chevalier+1980} found that in SNR non-radiative shocks the broad component is in Case A, while the narrow one lies in between Case A and Case B \citep[see model results in][]{Ghavamian+2001}.
In this work we always assume Case A for the broad-line component, and Case B for the narrow one.
The latter assumption is not fully justified in some cases, and could lead to overestimate to some
extent the intensity of the narrow-line component.

\section{Approximate analytic models}

As shown above, an accurate modelling requires a kinetic approach for the neutral components.
This has basically two consequences: that the system cannot be described by hydrodynamic equations; and that the microphysics parameters, like the reaction rates, are not constant, since the velocity distributions from which they are computed are non-thermal and change with position.

On the other hand, simplified models may be useful to quickly get reasonably accurate profiles, to effectively interpolate the (necessarily limited) number of cases treated in a fully numerical way, to allow in this way an optimization of the model parameters, and in general to match more effectively theory and data.
It should however be clear that the simplified models presented below are not intended to substitute the correct approach of the numerical simulations, but just to be useful interfaces between simulations and data.

\subsection{A 3-fluid model}

Here we present a simplified analysis, in which cold neutrals, hot neutrals, and protons are treated as three fluids: the first one moving at a velocity $\Vsh$ (where $\Vsh$ is the shock velocity), while the other two at $\Vsh/r$ (where $r$ is the compression factor).
The continuity equations, describing the spatial evolution of these three species, are then:
\begin{eqnarray}
\label{eq:fnevol}
\Vsh \frac{d}{dz}\left(f_n(z)\right)\!\!\!\!&=&\!\!\!\!-(\kp_{ce}+\kp_{i,n})f_n(z)f_{p}(z)	;	\\
\label{eq:fbevol}
\Vsh \frac{d}{dz}\left(f_b(z)/r\right)\!\!\!\!&=&\!\!\!\!\left(\kp_{ce}f_n(z)-\kp_{i,b}f_b(z)\right)f_{p}(z);		\\
\label{eq:fievol}
\Vsh \frac{d}{dz}\left(f_{p}(z)/r\right)\!\!\!\!&=&\!\!\!\!\left(\kp_{i,n}f_n(z)+\kp_{i,b}f_b(z)\right)f_{p}(z),	
\end{eqnarray}
where $f_n$, $f_b$, and $f_{p}$ are the densities of cold neutrals, hot neutrals and protons, normalized to the upstream gas density $n_0$, from which it follows that the length unit scales with $n_0^{-1}$
(in choosing the subscripts $n$ and $b$ we referred to the fact that the cold and hot neutral components are associated to the ``narrow'' and ``broad'' spectral components respectively).
The reaction rates included are $\kp_{ce}$ for the charge-exchange process, and respectively $\kp_{i,n}$ and $\kp_{i,b}$ for the ionization of cold neutrals, and of hot neutrals.
The fact that the backward flow of neutrals \citep[as from][]{paperI, Hester+1994} is not explicitly present in the above equations is justified by the fact that it is less important at large shock speeds.
On the other hand, the analytic solutions developed here will be used to fit results from kinetic models, in which the neutral return flux is correctly taken to account. In fact, as shown in Table~\ref{tab:physparams}, the best-fit solutions require a positive value of $f_b$, although very small, right at the shock. This value is smaller for increasing shock speed: we interpret this as an effect of the presence of a neutral precursor in the kinetic models.

A direct consequence of the above equations is the conservation of the total particle flux:
\begin{equation}
f_n(z) + \left[ f_b(z) + f_{p}(z) \right]/r=1.
\end{equation}
In general $\kp_{ce}$, $\kp_{i,n}$, $\kp_{i,b}$ and $r$ should change with space, but from here on we will assume all these quantities to be spatially constant.
Let us introduce $\kp_R=\kp_{ce}+\kp_{i,n}$ as a reference reaction rate, and scale all reaction rates with it:
\begin{eqnarray}
\kp_{ce}  &=& g_{i,n} \kp_R;	\\
\kp_{i,n} &=& (1-g_{i,n})\kp_R;		\\
\kp_{i,b} &=& (1-g_{i,b})\kp_R/r.
\end{eqnarray}
The newly defined quantities $g_{i,n}$ and $g_{i,b}$ are dimensionless constants, suitably chosen to simplify the following expressions.

Eqs.~(\ref{eq:fnevol})--(\ref{eq:fievol}) do not generally allow an analytic solution with respect to the variable $z$, but they do with respect to $f_n$, taken as the independent variable ($z$ then disappears, since it is not explicitly present in the equations).
The general solution is:
\begin{equation}
\frac{f_b(f_n)}{f_n}=\left(\frac{f_{b,0}}{f_{n,0}}+\frac{r\,g_{i,n}}{g_{i,b}}\right)\left(\frac{f_n}{f_{n,0}}\right)^{-g_{i,n}}
-\frac{r\,g_{i,n}}{g_{i,b}}.
\end{equation}
In the special case of a low neutral fraction, we can assume $f_{p}$ to be constant (and equal to $r$); in which case Eq.~(\ref{eq:fnevol}) can be easily solved giving:
\begin{equation}
f_n(z)=f_{n,0}\exp\left(-\frac{z}{h_R}\right),
\end{equation}
with $h_R=\Vsh/(r\kp_R n_0)$ being the reference scale length; and therefore:
\begin{equation}
\label{eq:fbovfnthree}
\frac{f_b(z)}{f_n(z)}=\left(\frac{f_{b,0}}{f_{n,0}}+\frac{r\,g_{i,n}}{g_{i,b}}\right)\exp\left(\frac{g_{i,n}z}{h_R}\right)
-\frac{r\,g_{i,n}}{g_{i,b}}.
\end{equation}
Once all density profiles have been derived, the emissivities in the narrow and broad line component can be approximated as:
\begin{eqnarray}
\label{eq:modeljn}
j_n(z)&=&\eps_{n}f_n(z)f_{p}(z)\,\kp_R\,n_0^2;	\\
\label{eq:modeljb}
j_b(z)&=&\eps_{b}f_b(z)f_{p}(z)\,\kp_R\,n_0^2,
\end{eqnarray}
where we have parametrized the physics of the emission with the parameters $\eps_{n}$ and $\eps_{b}$, which are dimensionless (since the emissivities are in $\U{ph\,cm^{-3}s^{-1}}$), and that for simplicity we also assume to be constant. The ratio of the line components is then:
\begin{eqnarray}
\!\!\!\!\!\!\!\!\!\!\!\!\frac{I_b(z)}{I_n(z)}\!\!\!&=&\!\!\!\frac{\eps_{b}}{\eps_{n}}\frac{f_b(z)}{f_n(z)}	\\
\!\!\!&=&\!\!\!\frac{\eps_{b}}{\eps_{n}}\left[\left(\frac{f_{b,0}}{f_{n,0}}+\frac{r\,g_{i,n}}{g_{i,b}}\right)\exp\left(\frac{g_{i,n}z}{h_R}\right)-\frac{r\,g_{i,n}}{g_{i,b}}\right]
\label{eq:IbovInmodel}
\end{eqnarray}
(hereafter we will use the symbols $j_{n,b}$ only for the emissivities, while $I_{n,b}$ represents the volume-integrated emission).

The spatially integrated emissions, scaled with their dimensional part, are then:
\begin{eqnarray}
\label{eq:integjn}
\frac{I_n}{n_0f_{n,u}\Vsh}&=&\eps_{n}\frac{f_{n,0}}{f_{n,u}};	\\
\label{eq:integjb}
\frac{I_b}{n_0f_{n,u}\Vsh}&=&\frac{\eps_{b} }{(1-g_{i,n})}\left(\frac{r\,g_{i,n}^2}{g_{i,b}}\frac{f_{n,0}}{f_{n,u}}+\frac{f_{b,0}}{f_{n,u}}\right),
\end{eqnarray}
where $f_{n,u}$ the upstream neutral fraction (which, in the presence of a precursor, may be slightly different from $f_{n,0}$, the value right at shock front). From the above equations one can derive the line ratio $I_b/I_n$.
When the shock structure is not resolved, these two quantities are the only ones that can be measured.

\begin{figure}
 \includegraphics[width=\columnwidth]{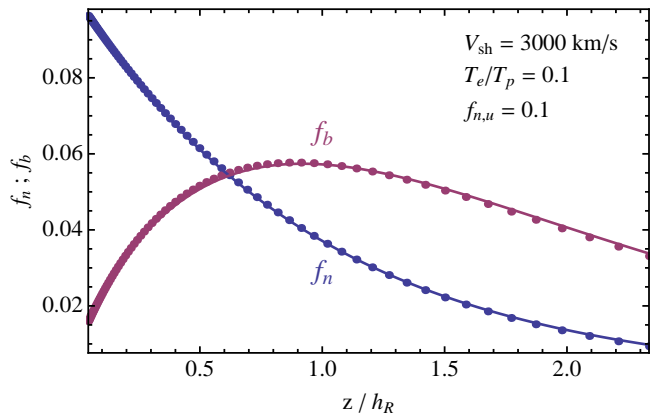}
 \caption{Comparison between numerical profiles (dots) of $f_n$ (blue) and $f_b$ (red) \citep[computed using the model presented in][and following papers]{paperI}, together with fits (solid lines) obtained using our approximate model. These profiles are for $\Vsh=3000\U{km\,s^{-1}}$, unit total density and $10\%$ neutral fraction upstream, and $T_e/T_p=0.1$.}
 \label{fig:EMFit3000rho}
\end{figure}

\begin{figure}
 \includegraphics[width=\columnwidth]{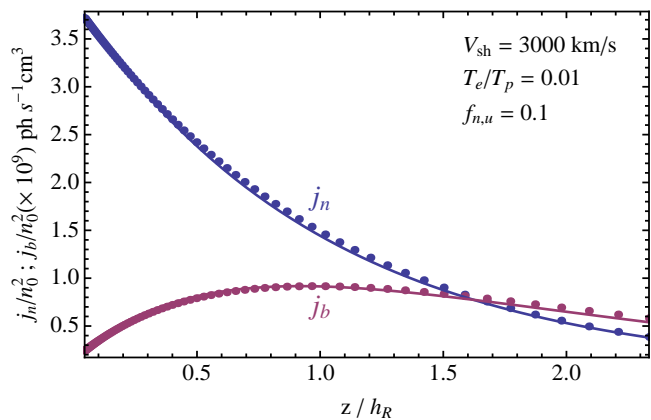}
 \caption{Same as Fig.~\ref{fig:EMFit3000rho}, but for the emissivity in the two line components.}
 \label{fig:EMFit3000jjj}
\end{figure}

We want to stress that the above approximated model will only be used to fit profiles calculated with our numerical simulations. Therefore, the assumptions behind it, like the constancy of several parameters and first of all the fluid-like treatment, are simply justified by how effectively it fits, with a limited number of ``physical-like'' parameters, the results of our kinetic models (as in Section~2).
For instance, the non-gaussianity of the velocity distribution of neutrals leads to some spatial changes for $r$ different from what we have assumed.
In addition, the use of Eq.~\ref{eq:modeljb} for modelling the emissivity in the broad component is in general not justified, because also charge-exchange reactions with slow and fast neutral lead to fast neutrals in excited states, and could be acceptable only in the limit of low neutral fractions.
We have found anyway reasonable fits for both the particle distributions and the emission in the two line components.
Table~\ref{tab:physparams} gives, in columns (4) to (10), the best-fit parameters to some kinetic models.
 Figs.~\ref{fig:EMFit3000rho} and \ref{fig:EMFit3000jjj} also show the quality of the fit for one of the models.

\begin{table*}
\caption{Best-fit parameters to our kinetic models (columns (4) to (10)); derived integrated $I_b/I_n$ ratio (11); same ratio, from more realistic simulated observations as in \citet{paperII} (12); correction factor, as from Eq.\ref{eq:IbovInobsA} (13). Symbols are defined in the text, and the compression factor $r$ is taken to be equal to 4.}
\begin{tabular}{ccccccccccccccc}
\hline
$\Vsh$&$T_e/T_p$&$f_{n,u}$&\hbox to0.2cm{\hfil}&$f_{n,0}$&$f_{b,0}$&$h_Rn_0$&$g_{i,n}$&\!\!\!$g_{i,b}/g_{i,n}$\!\!\!&$\eps_{n}$&$\eps_{b}$&\hbox to0.2cm{\hfil}&\!\!\!$I_b/I_n$\!\!\!&\!\!\!$I_b/I_n$\!\!\!&$\al$\\
${\rm km\,s^{-1}}$&&&&&&$\times10^{14}\,{\rm cm^{-2}}$&&&&&&\!\!\!present\!\!\!&\!\!\!M+12\!\!\!&\\
(1)&(2)&(3)&&(4)&(5)&(6)&(7)&(8)&(9)&(10)&&(11)&(12)&(13)\\
\hline
2500&0.01&0.10&&0.102&0.020&5.184&0.057&0.157&0.088&0.027&&0.52&0.72&0.087\\
2500&0.10&0.10&&0.102&0.021&5.504&0.044&0.097&0.079&0.026&&0.68&1.24&0.216\\
3000&0.01&0.10&&0.101&0.012&6.837&0.017&0.065&0.106&0.037&&0.42&0.52&0.061\\
3000&0.10&0.10&&0.101&0.012&7.452&0.010&0.028&0.097&0.040&&0.63&0.69&0.012\\
3500&0.01&0.10&&0.100&0.009&8.838&0.007&0.041&0.121&0.044&&0.29&0.31&0.021\\
3500&0.10&0.10&&0.100&0.006&9.708&0.018&0.069&0.113&0.047&&0.47&0.47&0.000\\
\hline
\end{tabular}
\label{tab:physparams}
\end{table*}

\begin{table*}
\caption{Derived parameters for the parametric model (columns (4) to (6)); same parameters, corrected with using Eqs.~\ref{eq:Ancorr}--\ref{eq:Bbcorr} (columns (7) to (9)).}
\begin{tabular}{ccccccccccc}
\hline
$\Vsh$&$T_e/T_p$&$f_{n,u}$&\hbox to0.3cm{\hfil}&$A_n$&\!\!\!$A_b/A_n$\!\!\!&\!\!\!$B_b/A_n$\!\!\!&\hbox to0.3cm{\hfil}&$A_n$&\!\!\!$A_b/A_n$\!\!\!&\!\!\!$B_b/A_n$\!\!\!\\
${\rm km\,s^{-1}}$&&&&&&&&(obs)&(obs)&(obs)\\
(1)&(2)&(3)&&(4)&(5)&(6)&&(7)&(8)&(9)\\
\hline
2500&0.01&0.10&&0.089&0.061&0.46&&0.081&0.162&0.50\\
2500&0.10&0.10&&0.080&0.068&0.61&&0.063&0.362&0.78\\
3000&0.01&0.10&&0.107&0.042&0.38&&0.100&0.110&0.40\\
3000&0.10&0.10&&0.097&0.050&0.58&&0.096&0.063&0.59\\
3500&0.01&0.10&&0.122&0.031&0.26&&0.119&0.053&0.26\\
3500&0.10&0.10&&0.113&0.027&0.44&&0.113&0.027&0.44\\
\hline
\end{tabular}
\label{tab:linearfit}
\end{table*}
\subsection{A parametric model}

Even the above simplified model may be too involved for a direct comparison with the data, due to possible degeneracies among some physical parameters.
We then prefer to use a parametrization closer to what is actually observed, in terms of polynomials of $z$ times an exponential function.

Here we discuss a very simple but non-trivial case, which already contains a number of features present in real cases.
Let us assume that the downstream profile of the emissivity in the narrow and broad-line component are respectively described by:
\begin{eqnarray}
\label{eq:jndef}
j_n(z)&=&n_0f_{n,u}\Vsh\frac{A_n}{h_R}\exp\left(-\frac{z}{h_R}\right);	\\
\label{eq:jbdef}
j_b(z)&=&n_0f_{n,u}\Vsh\left(\frac{A_b}{h_R}+\frac{B_bz}{h_R^2}\right)\exp\left(-\frac{z}{h_R}\right);
\end{eqnarray}
the former formula is equivalent to that of our 3-fluid model (Eq.~\ref{eq:modeljn}), in the limit of small neutral fractions; while the latter one, much simpler than that  (Eq.~\ref{eq:modeljb}) in the previous section, is anyway a good approximation.
In particular, the assumption of the same exponential length scale $h_R$ for both emissivities is justified by the fact that, in Eq.~\ref{eq:fbovfnthree}, the parameter $g_{i,n}$ is always very small, so that a linear expansion of the exponential factor is accurate till large $z/h_R$ values; but as shown below the most direct justification comes the fact that fits to the kinetic models are very good.

It is possible to relate the quantities $A_n$, $A_b$ and $B_b$ to the parameters used in the 3-fluid model described above.
From a comparison with a power expansion in $z$ of Eqs.~\ref{eq:modeljn} and \ref{eq:modeljb}, from our models one derives:
\begin{eqnarray}
A_{n}&=&\eps_{n}f_{n,0}/f_{n,u};	\\
A_{b}&=&\eps_{b}f_{b,0}/f_{n,u};	\\
B_{b}&=&g_{i,n}\left(\frac{f_{b,0}}{f_{n,u}}+\frac{r\,g_{i,n}}{g_{i,b}}\frac{f_{n,0}}{f_{n,u}}\right)\eps_{b}.
\end{eqnarray}
These three quantities are listed for all the models presented in Table~\ref{tab:linearfit}, at columns from (4) to (6).

\begin{figure}
 \includegraphics[width=\columnwidth]{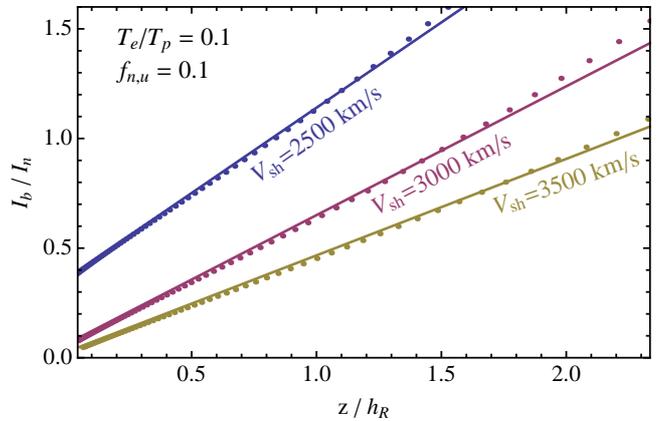}
 \caption{Profiles of $I_b/I_n$ for three models, respectively with $\Vsh=2500$, $3000$ and $3500\U{km\,s^{-1}}$ (and $f_{n,u}=0.1$, $T_e/T_p=0.1$), all corrected using Eq.~\ref{eq:IbovInobsA}. The dots are from the kinetic models (whose locations are very well reproduced by our 3-fluid model), while the lines show their linear approximations, provided by Eq.~\ref{eq:linearapprox}.}
 \label{fig:AlmostLinear}
\end{figure}
It can be seen that, within this approximation, the spatially integrated emissions divided by $(n_0f_{n,u}\Vsh)$ are $A_n$ and $A_b+B_b$ for the narrow and the broad component, respectively (see for comparison Eqs.~\ref{eq:integjn} and \ref{eq:integjb}).
So, their ratio is $I_b/I_n=(A_b+B_b)/A_n$.
Another consequence of this formulation is that the line components ratio along $z$ must follow a linear trend, namely:
\begin{equation}
\frac{j_b}{j_n}=\frac{A_b}{A_n}+\frac{B_b}{A_n}z.
\label{eq:linearapprox}
\end{equation}
This is valid only as a first approximation (see Eq.~\ref{eq:IbovInmodel} for a more accurate one), but it can be verified that, for the $z$ values of interest, it is a reasonably good approximation (see Fig.~\ref{fig:AlmostLinear}).

\begin{figure}
 \includegraphics[width=\columnwidth]{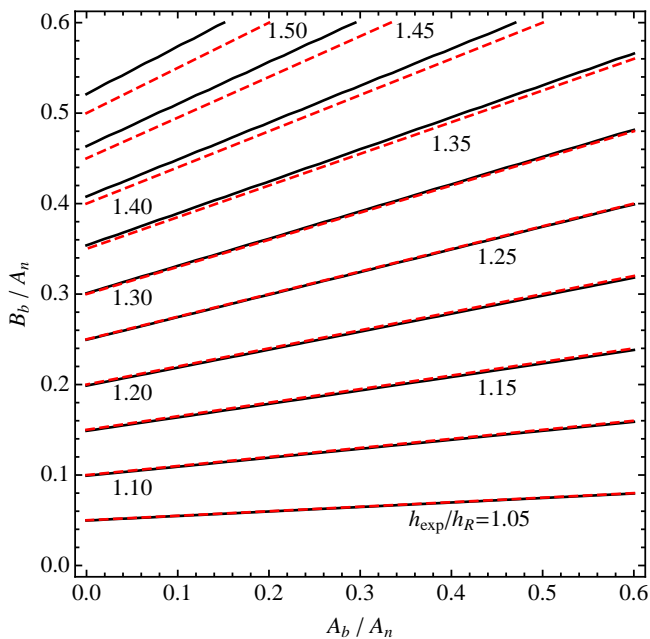}
 \caption{Contour plot giving the value of the ratio $\hexp/h_R$ as a function of $A_b/A_n$ and $B_b/A_n$. The black solid lines trace the results of exponential best fits to the profiles (down to 5\% of their peak values). The red dashed lines trace instead the approximated values, obtained using Eq.~\ref{eq:hexpovhRapprox}.}
 \label{fig:hexpovhRrelation}
\end{figure}
It is worthwhile to stress that when information on narrow and broad line components is not available separately, the quantities $A_b/A_n$ and $B_b/A_n$ cannot be inferred observationally.
One can proceed fitting the total line flux, in which case the simplest (and, as we shall see below, sufficiently accurate) fit is that using a pure exponential law, namely:
\begin{equation}
j_n+j_b=C\exp\left(-\frac{z}{\hexp}\right).
\label{eq:exponapprox}
\end{equation}
In this case, one then needs to relate the exponential length scale ($\hexp$) to the previously defined $h_R$.
Since the exact functional dependencies using $h_R$ and $\hexp$ are different, the best-fit relation would depend on the interval of $z$ used for the fit. Anyway, we have found that a simple and rather accurate relation is:
\begin{equation}
\frac{\hexp}{h_R}=1+\frac{B_b}{A_n+A_b}.
\label{eq:hexpovhRapprox}
\end{equation}
Fig.~\ref{fig:hexpovhRrelation} compares the results obtained with this formula and those of exponential fits to our parametric model.

\subsection{Matching the observed $I_b/I_n$ ratio}

The values of the integrated $I_b/I_n$ ratios, as derived from our models and listed in column (11) of Table~\ref{tab:physparams}, are always lower than those we had computed in \citet{paperII}, shown in Fig.~16 therein, and listed in column (12) of Table~\ref{tab:physparams}.
The reason of this discrepancy is that in \citet{paperII} the synthesized line profiles have been fitted with a two-component model in a very similar way to what is usually done in the analysis of actual observations (with a simulated instrumental spectral resolution $\sim150\U{km\,s^{-1}}$): this approach is feasible on spatially integrated models, but rather cumbersome and less reliable if one wants to apply it to each spatial step of our model.

Instead, what we dub here as ``narrow'' component is the composition of two different kinds of populations: the truly ``cold'' neutrals, namely those directly coming from the upstream and with thermal velocities $\sim10\U{km\,s^{-1}}$, and those originated instead from a charge exchange with a warm proton in the neutral precursor, having thermal velocities typically in the 100--300$\U{km\,s^{-1}}$ range; this latter population is the one emitting the ``intermediate component'', as described in our past papers, and  clearly detected in N103B by \citet{Ghavamian+2017} and in Tycho's SNR by \citet{Knezevic+2017}.

However, in all cases in which the quality of the observation is not good enough to detect the intermediate component, we expect that, after the line profile fit, a fraction of the flux of the intermediate component is partly ascribed to the narrow component, and partly to the broad component.

Since in the downstream the spatial behaviours of cold and intermediate neutrals are very similar, we assume that a spatially constant fraction $\al$ of the $I_n$, as calculated by the model, is contributing to the observed broad component, so that the observed $I_b/I_n$ ratio, in terms of the model fluxes, reads:
\begin{equation}
\label{eq:IbovInobsA}
\left.\frac{I_b}{I_n}\right|\rs{obs}=\frac{I_b/I_n+\al}{(1-\al)}.
\end{equation}
The last two columns of Table~\ref{tab:physparams} show the ratio between the integrated broad and narrow components, respectively, as it would be observed \cite[see Fig.~16 in][]{paperII}, and the value of $\al$ required for the correction.
The fact that the values of $\al$ decrease for increasing shock velocity is easily explained by a decrease, with increasing $\Vsh$, of the relative effectiveness of charge-exchange processes, so that less warm neutrals may cross back the shock, therefore forming a less prominent neutral precursor, and consequently a weaker intermediate component.

As for our parametric model, using Eq.~\ref{eq:IbovInobsA} the expected values for the observed quantities, as functions of the model ones, are:
\begin{eqnarray}
\label{eq:Ancorr}
A_{n,\mathrm{obs}}&=&(1-\al)A_{n};	\\
\label{eq:Abcorr}
A_{b,\mathrm{obs}}&=&A_{b}+\al A_{n};	\\
\label{eq:Bbcorr}
B_{b,\mathrm{obs}}&=&B_{b}.
\end{eqnarray}
These quantities are listed in Table~\ref{tab:linearfit}, at columns from (7) to (9).
From here on, when referring to the quantities $A_{n,\mathrm{obs}}$,  $A_{b,\mathrm{obs}}$, and $B_{b,\mathrm{obs}}$, for simplicity we will omit the suffix ``obs''.

\section{Analytic projected profiles}

In Eqs.~\ref{eq:jndef} and \ref{eq:jbdef} we have introduced the simplest non trivial way to model parametrically the emissivity profiles in the two line components.
On the other hand, virtually any profile could be approximated with arbitrary accuracy, by increasing the order of the polynomials (now respectively 0 and 1) in those formulae.
In this section we set the mathematical basis to model, for any intrinsic downstream profile, the actual observed profiles by taking into account projection effects.

\subsection{The limit of large curvature radii}

The general problem of connecting the downstream emissivity profiles to their transformation into observed surface brightness profiles, for a generically curved shock surface, is numerically complex and heavy.
Therefore, performing a best-fit analysis on data profiles may become a difficult task.

We present here an analytic treatment that can be used to considerably simplify this problem.
To do so, we assume that the curvature radii of the shock surface are always much larger than the projected shock distance  ($\dl z$; again oriented to the downstream) that we are considering: this allows us to neglect its component along the LOS ($\dl y$) when computing the distance of a point from the shock surface, which is equivalent to limiting to the first order all expansions in $z_r/\Rcurv$, where $z_r$ is the actual distance of the point from shock (positive downstream) and $\Rcurv$ is the local radius of curvature (positive for a convex curvature), namely:
\begin{equation}
z_r\simeq \dl z = z-\Zsh(y),
\end{equation}
where $\Zsh(y)$ describes the shock surface, with $y$ being the LOS coordinate.

The surface brightnesses of the narrow and broad components, $I_n(z)$ and $I_b(z)$ respectively, can be computed by integrating along the LOS the downstream emissivities $j_n(z)$ and $j_b(z)$:
\begin{equation}
I_{n,b}(z)=\int{j_{n,b}(z-\Zsh(y))\,dy},
\end{equation}
where the integration limits are derived by solving in $y$ the equation $z=\Zsh(y)$, and retaining only the downstream segment(s) along the LOS.

\subsection{Analysis of the constant-curvature case}

For a constant curvature, in the limit of large $\Rcurv$, one gets:
\begin{equation}
\label{eq:ZshlargeRcurv}
\Zsh(y)=\Rcurv\left(1-\sqrt{1-\frac{y^2}{\Rcurv^2}}\right)\simeq\frac{y^2}{2\Rcurv}
\end{equation}
where, for simplicity, we have chosen $y=0$ as the reference point along the LOS, and $\Zsh(0)=0$.
Note that $\Rcurv$ must be taken with its sign, and therefore $\Zsh\ge0$ in the convex case, while $\le0$ in the concave one.

The intersections with the shock surface are respectively $y_{1,2}=\pm\sqrt{2z\Rcurv}$.
This means that, in the convex case, the surface brightness is positive only for $z>0$, and the integration must be performed along the path between these intersections.
In the concave case, instead, for $z<0$ the integration must be performed in the two outer paths, while for $z>0$ the integration must be performed over all $y$ values.

Let us now assume that the emissivity profiles $j_{n,b}(z)$ can be reasonably well approximated as a linear combination of (a sufficiently small number of) terms;
\begin{equation}
\chi_m(z, h)=(z/h)^m\exp(-z/h).
\end{equation}
Here below we shall show that, in the case of a constant curvature, either positive or negative, each one of these terms allows an analytic solution; and, therefore, an analytic solution is also possible for any linear combination of them.

\begin{figure}
 \includegraphics[width=\columnwidth]{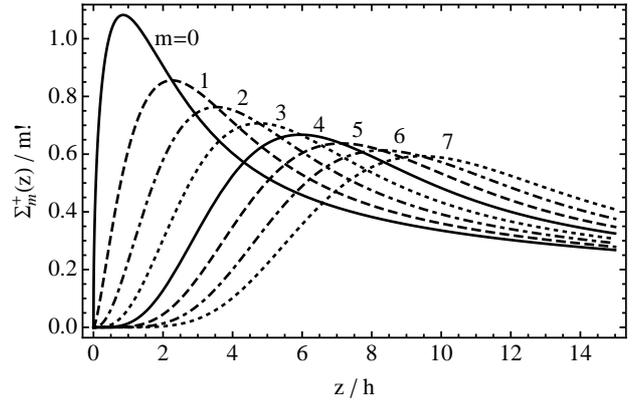}
 \caption{Profiles of $\Sg_m^{+,d}(x)$ (only projected downstream), scaled with $m!$ to give the same asymptotic solution at all values of $m$.}
 \label{fig:ProfsNconv}
\end{figure}

Let us first consider the case of a positive curvature (convex case).
In general it holds:
\begin{eqnarray}
\!\!\!\!\!\!\!\!\!\!\!\!\!\!\!\Sg_m^{+,d}(z)\!\!\!\!\!&=&\!\!\!\!\!\int_{y_1}^{y_2}{\!\!\!\!\chi_m\left(z-\frac{y^2}{2\Rcurv},h\right)\,dy}\nonumber\\
\label{eq:ConvexDSn}
\!\!\!\!\!&=&\!\!\!\!\!\Sg\rs{D}\left[P_m(z/h)\,2\FDawson\left(\!\sqrt{z/h}\,\right)-Q_m(z/h)\sqrt{z/h}\right],
\end{eqnarray}
where $\Sg\rs{D}=\sqrt{2h\Rcurv}$ is the dimensional scaling, namely an ``equivalent path length'', and $\FDawson(z)$ is the Dawson's integral, defined as:
\begin{equation}
\FDawson(z)=\exp(-z^2)\int_0^z{\exp(y^2)\,dy}=\frac{\sqrt{\pi}}{2}\exp(-z^2)\erfi(z),
\end{equation}
while the polynomials $P_m(z)$ and $Q_m(z)$ are given in Table~\ref{tab:FGcoefs} for some values of $m$.
In this case the surface brightness from the projected upstream region vanishes; while the profiles in the projected downstream are shown in Fig.~\ref{fig:ProfsNconv}.
\begin{table}
 \caption{Explicit forms of the first $P_m(z)$ and $Q_m(z)$ polynomials, which have been introduced in Eqs.\ from \ref{eq:ConvexDSn} to \ref{eq:ConcaveDSn}.}
 \label{tab:FGcoefs}
 \begin{tabular}{lll}
  \hline
$m$&$P_m(z)$&$Q_m(z)$\\[2pt] 
  \hline
0&$1$&$0$\\[2pt] 
1&$z+\frac{1}{2}$ &$1$ \\[2pt] 
2&$z^2+z+\frac{3}{4}$ &$z+\frac{3}{2}$ \\[2pt] 
3&$z^3+\frac{3}{2}z^2+\frac{9}{4}z+\frac{15}{8}$ &$z^2+2z+\frac{15}{4}$ \\[2pt] 
4&$z^4+2z^3+\frac{9}{2}z^2+\frac{15}{2}z+\frac{105}{16}$&$z^3+\frac{5}{2}z^2+\frac{25}{4}z+\frac{105}{8}$ \\
  \hline
 \end{tabular}
\end{table}

In an analogous way one can compute the case of a negative curvature (concave case).
Now the emission comes both from the projected upstream ($z<0$) and the projected downstream ($z>0$).

\begin{figure}
 \includegraphics[width=\columnwidth]{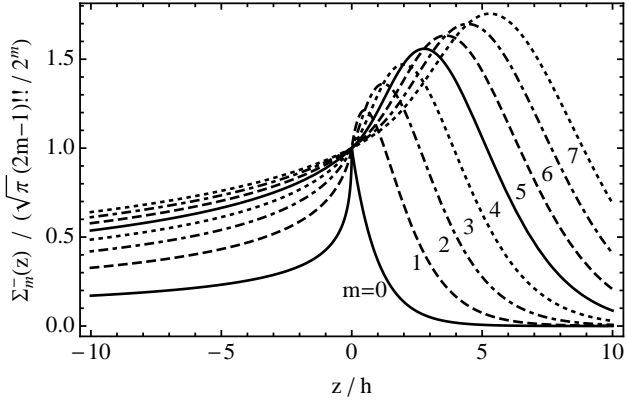}
  \caption{Profiles of $\Sg_m^{-}(z)$ (both projected upstream and downstream), scaled with $\sqrt{\pi}(2m-1)!!/2^m$ to normalize their value at $z=0$.}
 \label{fig:ProfsNconc}
\end{figure}

For $z<0$ we have:
\begin{eqnarray}
\!\!\!\!\!\Sg_m^{-,u}(z)\!\!\!\!\!&=&\!\!\!\!\!\int_{-\infty}^{y_1}{\!\!\!\!\chi_m\left(z+\frac{y^2}{2\Rcurv},h\right)\,dy}+
\int_{y_2}^{\infty}{\!\!\!\!\chi_m\left(z+\frac{y^2}{2\Rcurv},h\right)\,dy}\nonumber\\
&\!\!\!\!\!\!\!\!\!=&\!\!\!\!\!\!\!\!\!\Sg\rs{D}\left[P_m(z/h)\,2F\rs{X}\left(\sqrt{-z/h}\right)+Q_m(z/h)\sqrt{-z/h}\right],
\label{eq:ConcaveUSn}
\end{eqnarray}
where:
\begin{equation}
F\rs{X}(z)=\exp(z^2)\int_z^\infty{\exp(-y^2)\,dy}=\frac{\sqrt{\pi}}{2}\exp(z^2)\erfc(z),
\end{equation}
while the polynomials $P_m(z)$ and $Q_m(z)$ are the same as in the previous case.
Instead, for $z>0$ we simply have:
\begin{equation}
\Sg_m^{-,d}(z)\!=\!\!\!\int_{-\infty}^{\infty}{\!\!\!\!\!\!\chi_m\!\left(z+\frac{y^2}{2\Rcurv},h\right)\,dy}=\Sg\rs{D}P_m(z/h)\sqrt{\pi}e^{-z/h}\!.
\label{eq:ConcaveDSn}
\end{equation}
The global projected profiles are then obtained by combining $\Sg_m^{-,u}(z)$ for negative $z$ values and $\Sg_m^{-,d}(z)$ for positive $z$ values: the results are shown in Fig.~\ref{fig:ProfsNconc}.

The linear combination of the projected profiles, obtained in this way for this basis of functions, will allow us to treat in a rather simple way a wide range of cases.

\section{Spatially resolved profiles}

The solutions derived in the previous section are adequate to treat generally complex profiles in the shock transition zone.
However, with the limited quality of the observations available so far one could hardly go beyond the simplest non trivial level, which we have labelled as parametric model, and described by Eqs.~\ref{eq:jndef} and \ref{eq:jbdef}, so that
\begin{eqnarray}
\label{eq:Inphenomconvex}
\frac{j_n}{n_0f_{n,u}\Vsh}\!\!\!\!&=&\!\!\!\!\frac{2A_n}{h_R}\chi_0(z,h_R);	\\
\frac{j_b}{n_0f_{n,u}\Vsh}\!\!\!\!&=&\!\!\!\!\frac{2A_b}{h_R}\chi_0(z,h_R)+\frac{B_b}{h_R}\chi_1(z,h_R).
\end{eqnarray}
Within this framework we will discuss here some simple configurations.
At this stage we do not aim at a quantitative match of existing cases, but rather at outlining their qualitative behaviour.
However, we will use the case $A_b/A_n=0.063$ and $B_b/A_n=0.59$, for consistency with Section~7.1, where we will analyze a specific Balmer filament of SN~1006.
\begin{figure}
 \includegraphics[width=1.0\columnwidth,bb= 74 68 284 270, clip=true]{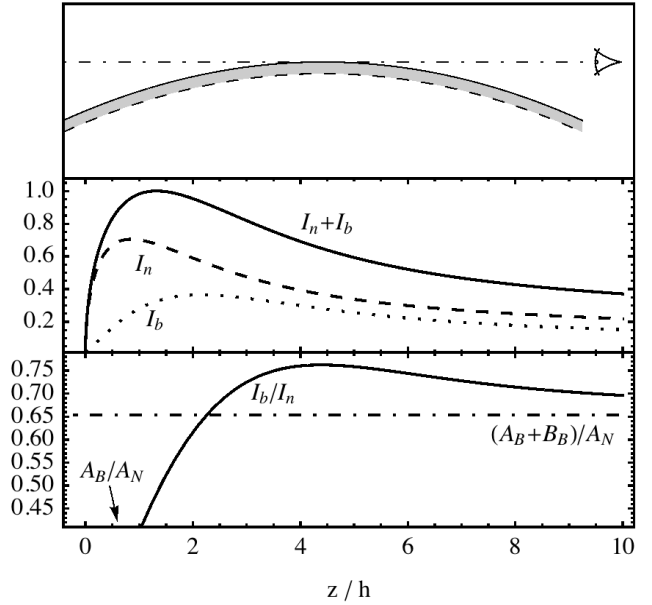}
  \caption{Upper panel: sketch of the geometry for a convex case with constant curvature, where the solid line represents the shock surface, the dashed line qualitatively represents the end of the transition zone, and the dot-dashed line the LOS corresponding to the projected limb ($z=0$). Mid panel: model intensity profiles, normalized to the maximum value of the total intensity. Lower panel: associated $I_b/I_n$ ratio; note the early increase, and the asymptotic convergence to the integrated ratio.}
 \label{fig:SampleConvex}
\end{figure}
\begin{figure}
 \includegraphics[width=1.0\columnwidth,bb= 74 68 284 270, clip=true]{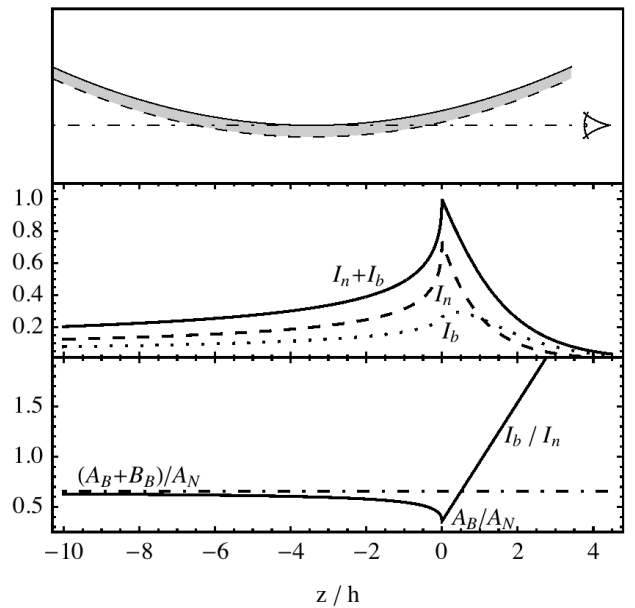}
  \caption{Same as Fig.~\ref{fig:SampleConvex}, for a concave case with constant curvature; note the upstream asymptotic value, the early decrease near the interface between projected upstream and projected downstream, and finally the linear divergence in the projected downstream.}
 \label{fig:SampleConcave}
\end{figure}
\begin{figure}
 \includegraphics[width=1.0\columnwidth,bb= 74 68 284 270, clip=true]{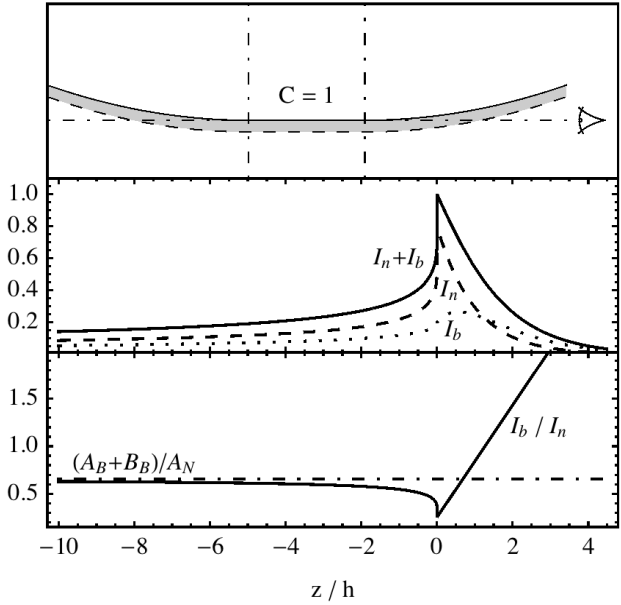}
  \caption{Same as Figs.~\ref{fig:SampleConvex} and \ref{fig:SampleConcave}, for a concave case with non-constant curvature, sketched by adding a straight line to the previous case. Here the $I_n+I_b$ profile is sensibly sharper, while the $I_b/I_n$ presents only minor changes.}
 \label{fig:SampleConcaveCC}
\end{figure}

\subsection{The case of a constant convex curvature}

In the convex case, one can use Eq.~\ref{eq:ConvexDSn} to derive:
\begin{eqnarray}
\label{eq:Inphenomconvex}
\frac{I_n}{n_0f_{n,u}\Vsh\Sg\rs{D}}\!\!\!\!&=&\!\!\!\!\frac{2A_n}{h_R}\FDawson\left(\sqrt{\frac{z}{h_R}}\right);	\\
\label{eq:Ibphenomconvex}
\frac{I_b}{n_0f_{n,u}\Vsh\Sg\rs{D}}\!\!\!\!&=&\!\!\!\!\frac{2A_b}{h_R}\FDawson\left(\sqrt{\frac{z}{h_R}}\right)	\nonumber\\
&&\!\!\!\!\!\!\!\!\!\!\!\!\!\!\!\!\!\!\!\!\!\!\!+\frac{B_b}{h_R}\left[\left(\frac{2z}{h_R}+1\right)\FDawson\left(\sqrt{\frac{z}{h_R}}\right)+\sqrt{\frac{z}{h_R}}\right]
\end{eqnarray}
in the projected downstream.
These curves are shown in Fig.~\ref{fig:SampleConvex}, together with $I_b/I_n$.

In the case of a constant curvature, the ratio between the surface brightness in the broad and narrow component ($I_b/I_n$) near the apex ($0<z\ll h_R$) increases as:
\begin{equation}
\frac{I_b}{I_n}\simeq\frac{A_b}{A_n}+\frac{2}{3}\frac{B_b}{A_n}\frac{z}{h_R};
\end{equation}
while, for large values of $z$, it reaches the asymptotic value $(A_b+B_b)/A_n$.

Therefore, already from a fit to the total emission one could verify the presence of such a geometry, estimating the scale length $\hexp$, and then $h_R$.
Observations of the line profile, with adequate angular resolution, allow one to trace the gradual increase of $I_b/I_n$, and to constrain separately $A_b/A_n$ and $B_b/A_n$.

\subsection{The case of a constant concave curvature}

For the concave case, we have:
\begin{eqnarray}
\frac{I_n}{n_0f_{n,u}\Vsh\Sg\rs{D}}\!\!\!\!&=&\!\!\!\!\frac{2A_n}{h_R}F\rs{X}\left(\sqrt{-\frac{z}{h_R}}\right);	\\
\frac{I_b}{n_0f_{n,u}\Vsh\Sg\rs{D}}\!\!\!\!&=&\!\!\!\!\frac{2A_b}{h_R}F\rs{X}\left(\sqrt{-\frac{z}{h_R}}\right)\nonumber\\
&&\!\!\!\!\!\!\!\!\!\!\!\!\!\!\!\!\!\!\!\!\!\!\!+\frac{B_b}{h_R}\left(\left(-\frac{2z}{h_R}+1\right)F\rs{X}\left(\sqrt{-\frac{z}{h_R}}\right)+\sqrt{-\frac{z}{h_R}}\right)
\end{eqnarray}
in the projected upstream, while:
\begin{eqnarray}
\label{eq:Inphenomconcave}
\frac{I_n}{n_0f_{n,u}\Vsh\Sg\rs{D}}\!\!\!\!&=&\!\!\!\!\frac{A_n}{h_R}\sqrt{\pi}e^{-z/h_R};	\\
\label{eq:Ibphenomconcave}
\frac{I_b}{n_0f_{n,u}\Vsh\Sg\rs{D}}\!\!\!\!&=&\!\!\!\!\left(\frac{A_b}{h_R}+\frac{B_b}{h_R}\left(\frac{z}{h_R}+\frac{1}{2}\right)\right)\sqrt{\pi}e^{-z/h_R}
\end{eqnarray}
in the projected downstream (as shown in Fig.~\ref{fig:SampleConcave}).

In the case of a constant curvature, $(A_b+B_b)/A_n$ is the $I_b/I_n$ asymptotic value at large negative values of $z$, while for $z$ approaching zero it decreases to $A_b/A_n$.
Finally, for the projected downstream regions, this simple model implies the following linear trend:
\begin{equation}
\frac{I_b}{I_n}=\frac{A_b}{A_n}+\frac{B_b}{A_n}\left(\frac{1}{2}+\frac{z}{h_R}\right).
\end{equation}
Then also in this case, with adequate spatial and spectral resolution, $A_b/A_n$, $B_b/A_n$ and $h_R$ could be extracted from the observations.

\subsection{A non-constant concave curvature}

\begin{figure}
 \includegraphics[width=1.0\columnwidth,bb= 82 68 284 210, clip=true]{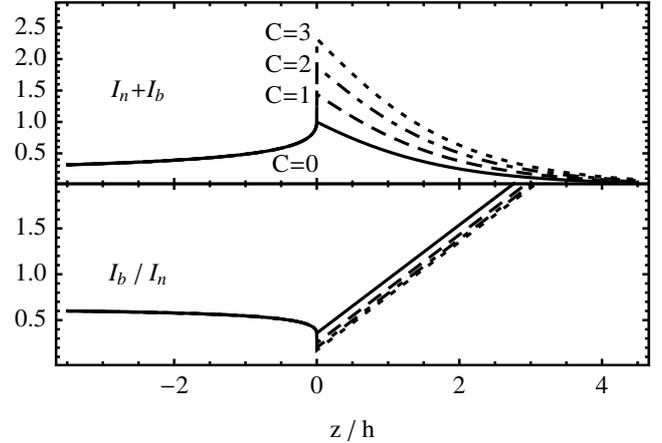}
  \caption{Upper panel: model total intensity profile in the constant curvature, concave case plus and edge-on straight component, for various values of $C$ ($C=0$ gives the same case as in Fig.~\ref{fig:SampleConcave}). Changing the value of $C$, the intensity profiles can change considerably. On the other hand (lower panel), the $I_b/I_n$ ratio is only slightly dependent on $C$.}
 \label{fig:SampleConcaveCCM}
\end{figure}
However, how we shall see below, rather often a concave model with constant curvature does not match the data well, so that a more sophisticated model is required.
Here we present a mild generalization of the constant curvature case, achieved by combining it with a 1-D model seen edge-on: this is similar to a sheet that does not follow a purely cylindrical geometry, but exhibits a tangent point to the LOS extending a significant distance along the LOS.
The results for this case are shown in Fig.~\ref{fig:SampleConcaveCCM}, for different values of the $C$ parameter: $C$ is a dimensionless parameter, which gives the length of the straight section in units of $\Sg\rs{D}$.
This still simple model does not match the data well, but shows some important facts: first, an extra edge-on layer, or more generally some deviations from a pure constant curvature which produce some wiggling near the tangent point, would enhance the magnitude of the peak compared to the surface brightness in the projected upstream; on the other hand, the profile of $I_b/I_n$ would undergo only minor changes.

\section{Application to a real case: SN 1006}

\subsection{Why SN~1006}

In the following we shall apply the methods derived so far to a real case, refining them whenever required.
The Balmer emission along the northwestern limb of SN 1006 represents for this an almost perfect case: observations with high spatial and/or spectral resolution, on which to test and use our models, are available; a shock velocity ($\simeq 3000\U{\kms}$) high enough to limit the role of charge-exchange processes (this simplifies the models and makes them more reliable); structures near the limb that appear rather well ordered; no evidence of efficient CR acceleration, which justifies neglecting the effect of CRs in the analysis.

With reference to the shock speed, \citet{Ghavamian+2001} measured a spectral width of $2290\pm80\U{km\,s^{-1}}$, from which for a low $T_e/T_p$ ratio they derived a shock speed of $2890\pm100\U{km\,s^{-1}}$.
A lower shock velocity, $\sim2500\U{km\,s^{-1}}$, follows instead from the model of \citet{paperIV}: for a more detailed and updated discussion see \citet{Raymond+2017}.
In the following analysis we shall fix $\Vsh$ to $3000\U{\kms}$.

With reference to CR acceleration, a low efficiency in the northwestern limb can be deduced from several results of observations:
1) the absence of non-thermal X-ray emission \citep{Bamba+2003,Winkler+2014}; 
2) a non detection of TeV \citep{Acero+2010} and GeV gamma-rays \citep{Xing+2016};
Moreover, from the analysis of Balmer emission, two additional pieces of information point towards a low CR acceleration efficiency:
3) the FWHM of the narrow  line is $\sim 21 \U{\kms}$ \citep{Sollerman+2003}, compatible with being produced by unperturbed ISM with $T= 10^4$ K, pointing toward the absence  of a CR precursor able to heat the upstream plasma;
4) an absence of evidence of any $\Hal$ precursor in front of the shock (R+07).

\subsection{Some recent high-quality observations}
\label{sec:summary}
In this section we summarize some results of two relevant papers, \citet[][hereafter R+07]{Raymond+2007} and \citet[][hereafter N+13]{Nikolic+2013}, in which Balmer emission from some areas along the northwestern limb of SN~1006 has been observed in great detail, and that contain a wealth of information, which will be used in the present work.
The former paper is based on a very deep HST/ACS image covering most of the $\Hal$ emission from that limb, and the superb spatial resolution of that image (a FWHM of about $0.13\arcsec$, corresponding to about $4\E{15}\dtwo\U{cm}$, where $\dtwo$ is the distance of SN~1006 in units of $2\U{kpc}$) allows one to probe scales shorter than the collisional mean free paths, and therefore to resolve the physical structure of the filaments.

The latter paper, instead, presents a map of a selected portion of the northwestern $\Hal$ limb of SN1006, roughly corresponding to ``Regions 26--29'' (as dubbed by R+07).
The instrument used, VIMOS in the IFU mode on the Very Large Telescope, allows both good spatial resolution (with a pixel size $0.67\arcsec$, corresponding to about $2.0\E{16}\dtwo\U{cm}$; and a typical seeing $\simeq1\arcsec$, namely $\simeq3\E{16}\dtwo\U{cm}$), and a spectral resolution $R\simeq2650$ (equivalent to about $110\U{km\,s^{-1}}$).

These newer data are complementary to those shown by R+07: now the spatial resolution is not as good as that of the HST; but the HST does not contain any information on the $\Hal$ line shape, while in VIMOS/VLT the spectral resolution is more than sufficient to measure the width of the broad line component ($W$), the intensity ratio between broad and narrow component ($I_b/I_n$), as well as the velocity shift of the centroid of the broad $\Hal$ component with respect to the narrow one ($\DlV$).

Let us first briefly report and discuss some conclusions in R+07, one of the main goals of which was to estimate the upstream medium density.
With the aim of analyzing the spatial structure of the $\Hal$ emission in the vicinity of the shock front, R+07 for the first time also discussed how the geometrical structure of the emitting region may affect the observations.
The approach of R+07 was to approximate a ripple in the shock surface as a concave portion of a cylindrical surface, and to try in this way a fit to the projected profile of a filament.
Within this framework, and with the help of a model for the physics in the downstream of the shock, the length scale of the inner part of the profile has been used to infer the total gas density; while the scale length of the outer part of the filament's profile, being mostly related to the curvature radius of the cylinder.
In turn, since the curvature radius is related to the effective emission length along the LOS, combining this information with a photometric measurement of the surface brightness R+07 also infer the gas neutral density (with the obvious constraint that it cannot be larger than the total density).
Therefore, from a combined fit to filament brightness and spatial profile one could aim at estimating independently curvature radius, total density and neutral fraction.

While this method is very powerful, the results presented in that paper were not definitive, in the sense that no combination of parameters was found, allowing for an exact match of the shape of the radial profiles (respectively ``Position 10'' and ``Position 28'', in that paper).
In addition, in both cases better fits have been obtained only with very small values for the curvature radius: about $10^{17}\U{cm}$, namely 100 times smaller than the observed length of the filament itself. In order to justify this large discrepancy between the two scale lengths the authors suggested the presence of a magnetic field oriented near the plane of the sky, as the reason for a strongly anisotropic pattern for the ripples.

In addition to all this, R+07 for the first time mentioned two effects, which in the following we will find to be very important: i. the possibility of more than one tangency point along the LOS, to explain the profile without requiring a too small curvature radius; ii. the possibility of bulk velocity contributions to the measured broad line width, if more than one layer intercepts the LOS. 

The second paper, namely N+13, presented a number of observed effects, unexpected before then, that surely deserve an explanation.
First of all, it showed clear evidence of significant spatial variations in $W$ (of order 10 to 20\% across the limb) and $I_b/I_n$ (ranging from $\sim0.4$ to $\sim1.6$), over just a few arcsec on the plane of the sky.
In particular, both spatial variations of $W$ and $I_b/I_n$ have rather distinctive behaviours across the rim: $W$ reaches larger values outwards of the bright filament while $I_b/I_n$ stays rather constant (with typical values of 0.7--0.8) outwards of that filament, but then shows strong variations right across the filament, reaching there both its lowest and its highest values within the field (see Fig.~\ref{fig:NikolicProfiles} for an overall view of all these trends).

The authors concluded that for the observed spatial variations one cannot simply invoke density variations, because variations up to 40\% would be required on small length scales (only tens of atomic mean free paths), and this would not be compatible with the smoothness of the shock observed even on much larger scales; instead, they proposed that these variations arise from the microphysics and that, in particular, the presence in some locations of very low values of the intensity ratio could motivate the need to include in the models suprathermal particles and CRs.

Another clearly observed effect is the velocity shifts between broad and narrow component, with values ranging from about $-300\U{\kms}$ to $+300\U{\kms}$: these shifts have been interpreted in terms of slightly different orientations of the local shock front, deviating in half of the cases by less than $\sim2$ degrees from the pure edge-on orientation (see their Table S1).
These estimates were based on the assumption of a single emitting layer along the LOS; while as we shall see below a different interpretation may be more plausible.

In the following, we shall adapt and generalize some of the concepts introduced by R+07, and we will show how effectively one can naturally explain, just in terms of geometrical effects, most of the phenomena mentioned above.

\begin{figure}
 \includegraphics[width=\columnwidth]{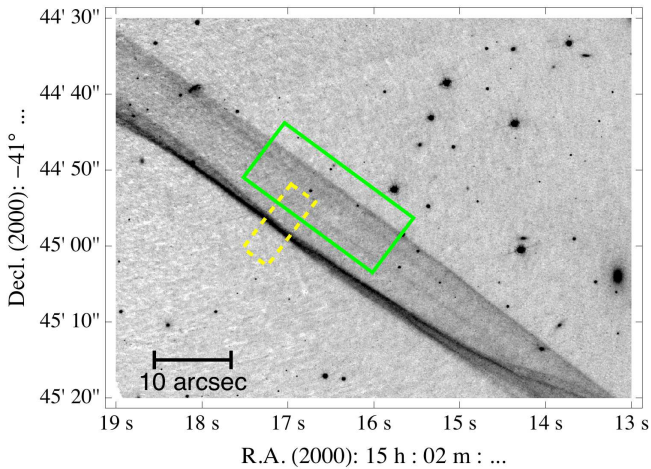}
 \caption{Locations of the two HST fields used for the analysis of the outer limb (green, solid-line rectangle), and the inner filament (yellow, dashed-line rectangle). A broader area has been chosen for the outer limb, in order to increase the signal-to-noise ratio, while the analyzed portion of the inner filament is smaller and limited to the region where the filament is sharper.}
 \label{fig:FieldwithAstrometry}
\end{figure}
\section{Re-analysis of Raymond et al. data}

In order to study in detail the filament's profile, and to extract further information from it, a higher spatial resolution is required, which in this case is provided by the HST.
The field considered here contains the area labelled as ``Regions 26--29" in  R+07 (see Fig.~\ref{fig:FieldwithAstrometry}).
In this zone the Balmer emission shows the following spatial structure: a stripe of about 10~arcsec width and, on its inner side, a much brighter and well defined filament, with a width of $\sim1$~arcsec.
Since in the following we analyze independently the two sectors, we use two different techniques for the data handling.

For the outer boundary we aim at reaching very low surface brightnesses.
For this reason we have chosen a wider area (shown in the figure by a green, solid line rectangle), and we have removed the regions containing stars.
As for the inner filament, instead, we have chosen a narrower area (outlined in the figure by a yellow, dashed-line rectangle, and rather close to Position 28, in R+07) to minimize blurring effects on the filament profile.

In all the fits shown below we will assume $A_b/A_n=0.063$ and $B_b/A_n=0.59$, as in Table~\ref{tab:linearfit}, corresponding to $\Vsh=3000\U{km\,s^{-1}}$, $T_e/T_p=0.1$, and $f_{n,u}=0.1$.
We will check the consistency of the data with this assumption {\it a posteriori}.

\subsection{Fitting the outer projected boundary}  \label{sec:out_boundary}
\begin{figure}
 \includegraphics[width=1.0\columnwidth,bb=5 5 355 230, clip=true]{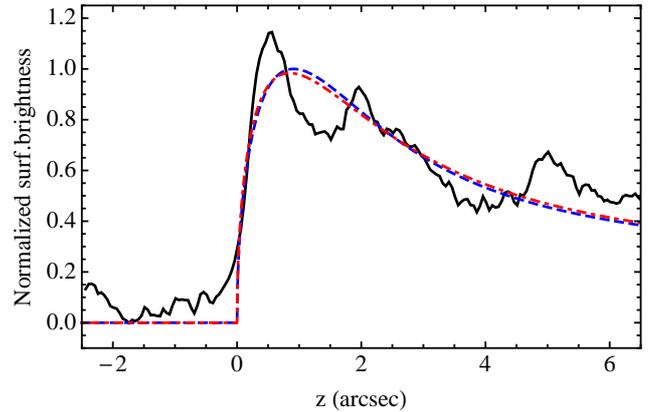}
 \caption{Comparison of the total intensity profile of the outer edge of the emission region, together with a fit obtained by using our parametric model (blue, dashed line) and with a fit using a simple exponential profile (red, dot-dashed line): note that the exponential fit gives already a reasonably good profile.}
 \label{fig:FitOuterLimb}
\end{figure}
Fig.~\ref{fig:FitOuterLimb} shows the intensity profile of the outer edge together with the fit results.
One may notice that the profile shows some limb brightening, which can be nicely fitted by a model with positive curvature.
We interpret residual oscillations in the observed profile as the effect of secondary, low amplitude ripples.

For one of the fits (blue dashed line) we have used our parametric model (sum of the Eqs.~\ref{eq:Inphenomconvex} and \ref{eq:Ibphenomconvex}), obtaining a length scale $h_R=0.69$~arcsec, which corresponds to an ambient density $n_0=0.033\,\dtwo^{-1}\U{cm^{-3}}$.

Also the result of a pure exponential profile (Eq.~\ref{eq:exponapprox}) is shown (red dot-dashed line), for which we have derived a best-fit value $\hexp=0.99$~arcsec that, using Eq.~ \ref{eq:hexpovhRapprox}, translates into $h_R=0.64$~arcsec: this, together with the almost coincidence of the two fitting curves, shows that a fit with an exponential profile is accurate enough.
Please notice that our fits also include secondary parameters such as the background level, the intensity scale, and the offset position of the limb.

Differently from the approach by R+07, here we cannot use the photometry to estimate the ambient density, because we do not have any way to estimate independently the path length along the LOS. Instead, by taking our previously estimated values for $n_0$ and $h_R$ and using Eqs.~\ref{eq:Inphenomconvex} and \ref{eq:Ibphenomconvex}, we can infer the ambient neutral fraction $f_{n,u}$, as a function of the local $\Rcurv$ along the LOS.
The normalized surface brightness profile shown in Fig.~\ref{fig:FitOuterLimb} can be converted into $\U{ph\,cm^{-2}s^{-1}arcsec^{-2}}$ multiplying it by a factor $2.69\E{-5}$. In this way we derive:
\begin{equation}
f_{n,u}=0.23\,(\Rcurv /1\U{arcsec})^{-1/2}.
\end{equation}
If along the LOS the curvature is similar to those seen on the plane of the sky, of order 50--500~arcsec, a neutral fraction of order 0.01--0.03 would be inferred, namely smaller than typically assumed.

\subsection{The effect of ripples on the shock surface}
\begin{figure}
 \includegraphics[width=1.0\columnwidth,bb=35 0 330 405, clip=true]{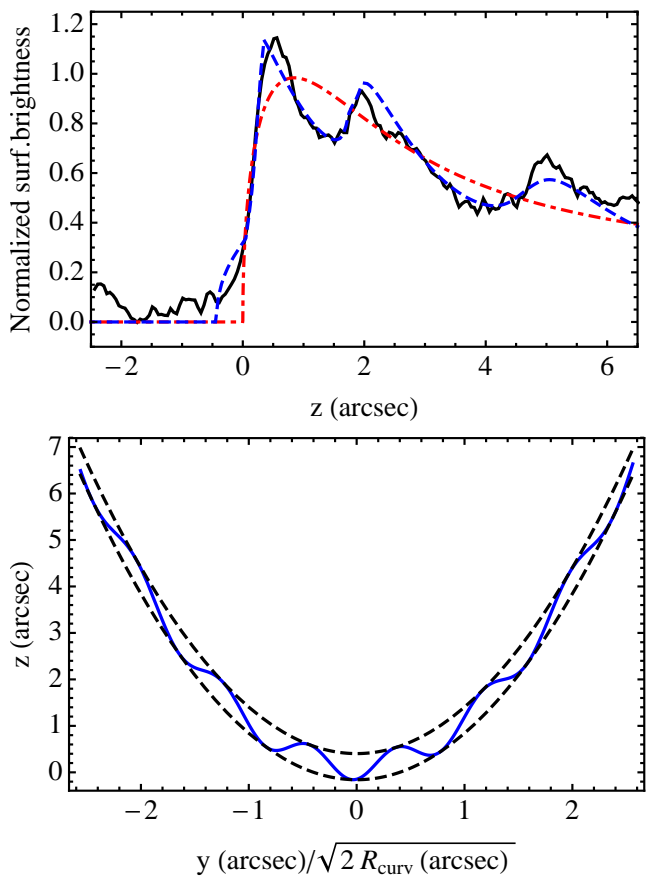}
 \caption{Results of the best-fit for a shock surface with sinusoidal ripples. The upper panel shows the new fit (blue dashed line), compared to the data (black solid line) and to the exponential fit (red dotted line), as already shown in Fig.~\ref{fig:FitOuterLimb}. The lower panel shows instead the shape of the shock surface (blue solid line), together with its upper and lower boundary (black dashed line) to make the oscillation more visible. Note that the horizontal and vertical scales are not the same: for the real shape the horizontal scale must be considerably stretched.}
 \label{fig:FitExptoouterlimbplusoscinshape}
\end{figure}
The fit in Fig.~\ref{fig:FitOuterLimb} actually shows a series of oscillations that are not reproduced by the smoother fit.
Therefore one may wonder whether, rather than the best-fit scale length $h_R$ (or alternatively $\hexp$ for an exponential fit), one sees instead the composition of structures with much smaller scale lengths (and therefore associated to much higher ambient densities).
In principle one may find an infinite number of mathematically valid solutions, which involve small-scale changes of the shock surface profile, and/or the ambient medium density, and/or the upstream neutral fraction.  The problem is how to match the observed radial profile with spatial distributions that are not clearly ``ad hoc'', and that are described by a rather small number of free parameters. 
In this section we will focus on the possibility that what is observed can be simply explained as a geometrical effect, due to the presence of small amplitude ripples.

For this, let us generalize Eq.~\ref{eq:ZshlargeRcurv} with the addition of a sinusoidal modulation:
\begin{equation}
\Zsh(y)=\frac{y^2}{2\Rcurv}+H\cos(k y+\phi)+Z\rs{offs}
\end{equation}
(the a small $Z\rs{offs}$ may be required for optimize the alignment to the data).
Note that also here we use the limit of large $\Rcurv$ values, in which case we cannot extract from the data information on $y$, $k$, $\Rcurv$ separately, but only on $\tilde y=y\sqrt{2\Rcurv}$ and $\tilde k=k/\sqrt{2\Rcurv}$.

We have then computed a least-square fitting to the observed profile of the outer limb, using the formula above for the profile of the shock surface, plus an exponential trend for the downstream emissivity profile.
The best fit parameters are:
\begin{equation}
\begin{tabular}{lll}
\!\!\!$\hexp=1.09\U{arcsec}$;&\!\!\!\!\!\!$H=0.24\U{arcsec}$;&\\
\!\!\!$\tilde k=8.0\U{arcsec^{1/2}}$;&\!\!\!\!\!\!$\phi=201\U{degree}$,&\!\!\!\!\!\!$Z\rs{offs}=-0.1\U{arcsec}$,
\end{tabular}
\end{equation}
and the resulting fit is shown in Fig.~\ref{fig:FitExptoouterlimbplusoscinshape}.
The quality of the fit is in a sense surprising, because in the reality one would not expect a pure sinusoidal behaviour, but rather random oscillations with some power spectrum. 
The information that one could infer from the best-fit parameters is the following: the value of $\hexp$ is consistent within 10\% with that derived in the previous section,without including oscillations, and therefore it must be taken as a rather robust result; the required amplitude of the oscillations ($H$) is very small; the wavelength of the oscillations $\lmb=2\pi/k$ is unknown, but can be rather long and scales as $\Rcurv^{1/2}$.

A further clue in favour of a purely geometrical effect can be derived from an inspection of the map in Fig.~\ref{fig:FieldwithAstrometry}. One may recognize the main peaks in the profile as stripes in the map, oriented ``almost'' parallel to the outer boundary. Incidentally, the fact that these features look almost parallel to the limb does not necessarily imply that the oscillations have a preferential orientation with respect to us, but simply that
the value of $\Rcurv$ is large (say, of order $10^3\U{arcsec}$).

\begin{figure}
 \includegraphics[width=1.0\columnwidth,bb=5 0 355 235, clip=true]{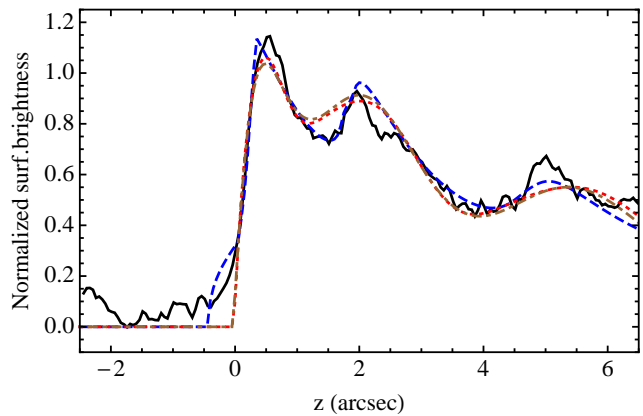}
 \caption{Comparison of best fits to the radial profile, by using respectively: a. oscillations in shape from the constant-curvature case (blue dashed curve); b. oscillations in density (red dotted color); c. oscillations in neutral fraction (brown dot-dashed curve). It is apparent that the first fit is slightly more accurate, and in particular that it is more effective in producing sharper secondary peaks.}
 \label{fig:CompareFitExptoouterlimbplusosc}
\end{figure}
We have also attempted analogous best fits, by first assuming a sinusoidal oscillation (of about $\pm60\%$) of only the ambient density, and then only of the neutral fraction (by an amount of $\pm80\%$). The results are shown in Fig.~\ref{fig:CompareFitExptoouterlimbplusosc}: in either case the secondary peaks look smoother that those obtained by perturbing the shock surface.
In fact, the data show some peaks, like for instance that one at $z\simeq2\U{arcsec}$, which are even sharper than the fit in Fig.~\ref{fig:FitExptoouterlimbplusoscinshape}.
Of course, a better fit could be obtained by allowing a more complex power spectrum than a simply monochromatic one, but this would require further free parameters and is beyond the scope of the present analysis.

\subsection{On the validity of the derived length scale}

Since the density estimate that we have obtained is significantly smaller than most of the previous estimated in the literature, it is worth discussing under which conditions our approach may overestimate the length scale $h$, therefore underestimating $n_0$.

To this purpose let us consider a limit case, in which a layer with vanishing $h$ mimics the case of an infinite $h$. This may occur, for instance, if the layer has exactly a triangular shape, with its vertex at the projected position of the outer edge: in this case we would see a sharp edge with a constant surface brightness inside.
However, it would be sufficient to have a slightly roundish vertex to get a very sharp peak of surface brightness near the projected limb, which then flattens to a low surface brightness value when the two flat shoulders are reached.

Since it is very unlikely (although possible) having such a sharp vertex, in the presence of a small value of $h$ one should likely see a very sharp peak followed by a stripe with a surface brightnesses much lower than those predicted by the constant curvature model (as from Fig.~\ref{fig:SampleConvex}).
Anyway, by a suitable combination of many such components with different offsets, one could manage to reproduce also the general profile as shown in Fig.~\ref{fig:FitOuterLimb}. But, in our opinion, it is not justified to assume such a fine tuned model to reproduce a general trend that is instead naturally reproduced by a constant curvature model.

To conclude, even if our estimated value $h_R=0.64\U{arcsec}$ is formally just an upper limit, we believe that it is a rather reliable estimate.

\subsection{Fitting the inner bright filament}

\begin{figure}
 \includegraphics[width=\columnwidth,bb= 20 5 340 365, clip=true]{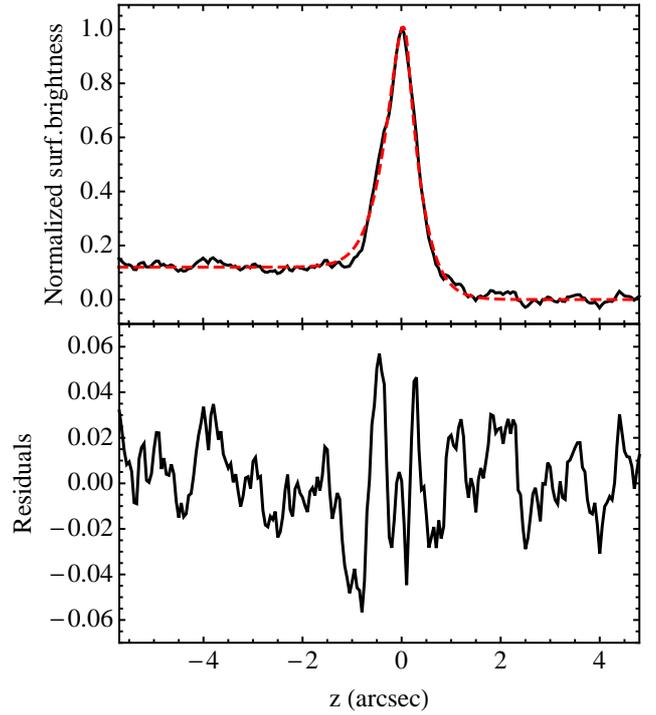}
 \caption{The upper panel shows the total line intensity profile of the inner filament, normalized to its maximum value (black line - the origin of the coordinate is set at the peak position), together with our best fit profile (red dashed line). The lower panel shows instead the fit residuals.}
 \label{fig:Fittoinnerlimb}
\end{figure}

The total intensity profile of the inner filament is shown in  Fig.~\ref{fig:Fittoinnerlimb}): as we have mentioned before, a constant concave curvature does not allow one to adequately describe its structure.
Here we present a method to fit its structure in detail, and to extract another independent estimate of the ambient density.

In order to perform this task, let us release our former assumptions about the shape of the shock front, and focus only on the distribution on the sky of the shock front positions, ${\cal P}(z)$.
Assuming that the emission profile across the transition zone ($j_t(z)=j_n(z)+j_b(z)$) is the same for all positions along the shock, the observed profile can be written as the convolution ${\cal P}(z)\ast j_t(z)$.
The inverse problem seems in principle unsolvable, because from one function (the observed profile), we aim at extracting two.

To allow a treatment of this problem we then proceed with two further assumptions.
First, that $j_t(z)$ follows our parametric model:
\begin{equation}
j_t(z)=\frac{1+(A_b/A_n)+(B_b/A_n)(z/h_R)}{1+(A_b/A_n)+(B_b/A_n)}\frac{\exp(-z/h_R)}{h_R},
\end{equation}
(scaled such that $\int_0^\infty j_t(z)\,dz=1$), where we assume $A_b/A_n$ and $B_b/A_n$ to be known, while leaving only $h_R$ as a free parameter.
The other assumption is that the ${\cal P}(z)$ distribution vanishes at all $z>\zmax$: this constraint retains the idea that the overall shape of the shock surface is concave there, and we shall see how adding this assumption a solution may be obtained.
One may find in fact that the profile in the projected downstream depends on ${\cal P}(z)$ through the following expression:
\begin{equation}
\label{eq:ItDSprofile}
I_{t,D}(z)=\left(C_{0,D}+C_{1,D}\frac{z-\zmax}{h_R}\right)\exp\left(-\frac{z-\zmax}{h_R}\right),
\end{equation}
where:
\begin{eqnarray}
C_{0,D}&=&\frac{(1+A_b/A_n){\cal I}_{(0)}-(B_b/A_n){\cal I}_{(1)}}{1+(A_b/A_n)+(B_b/A_n)};\\
C_{1,D}&=&\frac{(B_b /A_n){\cal I}_{(0)}}{1+(A_b/A_n)+(B_b/A_n)},
\end{eqnarray}
with:
\begin{equation}
{\cal I}_{(n)}=\int_{-\infty}^{\zmax}{\left(\frac{z'}{h_R}\right)^n\exp\left(\frac{z'}{h_R}\right){\cal P}(z')\,dz'};
\end{equation}
unfortunately, ${\cal I}_{(0)}$ and ${\cal I}_{(1)}$ can be evaluated only if the profile of ${\cal P}(z)$ is known.
In the projected upstream the evaluation is similar but more complex, since the upper limit of the integrals is now $z$, therefore changing with position.

\begin{figure}
 \includegraphics[width=\columnwidth]{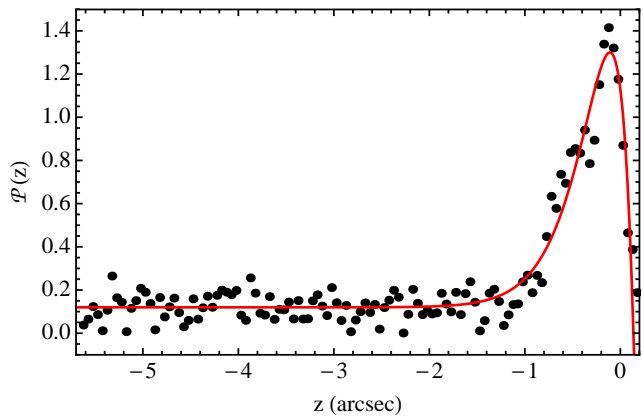}
 \caption{Comparison of our model ${\cal P}(z)$, as from Eq.~\ref{eq:emittersshape} (red line), and the distribution computed to match exactly the data (black dots).}
 \label{fig:CompareDistributionsofemitters}
\end{figure}

The problem can be further simplified making a guess for the function ${\cal P}(z)$.
We have tried the following form:
\begin{equation}
 \label{eq:emittersshape}
{\cal P}(z)=\left(a+b(z-\zmax)\right)\exp\left(\frac{z-\zmax}{h}\right)+c,\quad\hbox{for}\;z<\zmax
\end{equation}
(which will be justified {\it a posteriori}), where the best values for $a$, $b$, $c$ and $h$ have to be evaluated.
The choice of this profile is purely phenomenological, and has been inspired by the data shape.
It can be shown that in the projected upstream:
\begin{equation}
\label{eq:ItUSprofile}
I_{t,U}(z)=\left(C_{0,U}+C_{1,U}\frac{z-\zmax}{h_R}\right)\exp\left(\frac{z-\zmax}{h}\right)+c.
\end{equation}
The quantities $C_{0,U}$, $C_{1,U}$, $C_{0,D}$, and $C_{1,D}$ can be written as functions of $a$,$b$, $c$, $h$ and $h_R$ (plus of course of $A_b/A_n$ and $A_b/A_n$, which here have been given {\it a priori}).
By least-square fitting at the same time to the upstream and downstream data, and leaving also free the quantity $\zmax$, we finally obtain:
\begin{eqnarray}
&&\zmax=0.14\U{arcsec};\;h_R=0.24\U{arcsec}\nonumber\\
&&\!\!\!\!\!\!\!\!\!\!\!\!\!\!\!\!a=-0.18;\;b=-14.43;\;c=0.12;\;h=0.24\U{arcsec},
\label{eq:bestfitInnerFilament}
\end{eqnarray}
implying an ambient density $n_0=0.095\,\dtwo\U{cm^{-3}}$.
As shown by Fig.~\ref{fig:Fittoinnerlimb}, the fit to the filament's profile is quite good: this already can be taken as a proof of the validity of the shape introduced with Eq.~\ref{eq:emittersshape}.

An independent test of the goodness of the function ${\cal P}(z)$ can be performed comparing the fitted expression with a ``brute-force'' deconvolution, obtained using a number of parameters equal to the number of data points (therefore without any treatment of the errors): these are shown in  Fig.~\ref{fig:CompareDistributionsofemitters}, and apart from the scatter the profiles look virtually identical.

The most relevant result concerns the value of $h_R$, which is almost $3$ times smaller than than previously derived from the outer part of the external filament: this implies that at the location of the brightest filament the shock is really moving through a medium that is $\sim3$ times denser than in the outer edge, provided that the other physical parameters do not change significantly.
This is a first evidence of density changes, even if such density changes alone cannot be responsible for the huge variations observed in surface brightness.

Our assumption of the existence of a given $\zmax$ is formally incorrect, in the sense that shock is a closed surface, so that sooner or later its projection should turn towards the center of the SNR.
Our assumption therefore is equivalent to assume that, before this happens, the face-on surface brightness must almost vanish.
The fact that, as it will be shown below, the observed profile of $I_b/I_n$ in the projected downstream presents a steady increase of the $I_b/I_n$ ratio as long as some emission is detected, namely a behaviour similar to that in Fig.~\ref{fig:SampleConcave}, seems to justify the correctness of our assumption.

Now, the normalized surface brightness profile shown in Fig.~\ref{fig:Fittoinnerlimb} can be converted into $\U{ph\,cm^{-2}s^{-1}arcsec^{-2}}$ multiplying it by a factor $1.17\E{-4}$. In this way we derive:
\begin{equation}
f_{n,u}=1.7
\,(\lmbLOS/1\U{arcsec})^{-1},
\end{equation}
where $\lmbLOS$ is the effective length of the emission region along the LOS (again expressed in arcsec, to allow a direct comparison with the size of the structures seen on the map).
For instance,  an upstream neutral fraction 0.01--0.03 would imply $\lmbLOS$ to be in the range 60--170~arcsec corresponding to about 0.6--1.7~pc.

\subsection{Multiplicity of the brightest filaments}

In the previous section we have shown that the assumption of a rippled surface, rather than one with a constant curvature, can successfully reproduce the available data without assuming local curvatures along the LOS much smaller than those measurable on the plane of the sky, as done by R+07.

The only remaining issue is if the assumption of ripples, i.e.\ of multiple shock intersections along the LOS, should be considered or not as a ``fine-tuning''.
The purpose of this section is to justify the idea that multiple intersections occur quite naturally, once one has selected from the whole field of view only those locations with the brightest (projected) filaments.

\begin{figure}
 \includegraphics[width=1.0\columnwidth, clip=true]{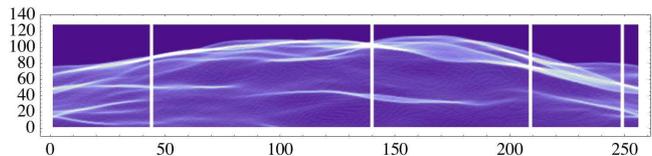}
  \caption{Simulated map, obtained by a single realization of a spherical sheet with random radial fluctuations (see text for details) and shown in projection. The four vertical stripes have been chosen among those containing the brightest portions of filaments, and the profiles for them are displayed in Fig.~\ref{fig:LimbTM_gridofslits}. 
The coordinate units, being arbitrary, are expressed here just in pixels.}
\label{fig:LimbTM_obsmapslits}
\end{figure}
\begin{figure}
 \includegraphics[width=1.0\columnwidth, clip=true]{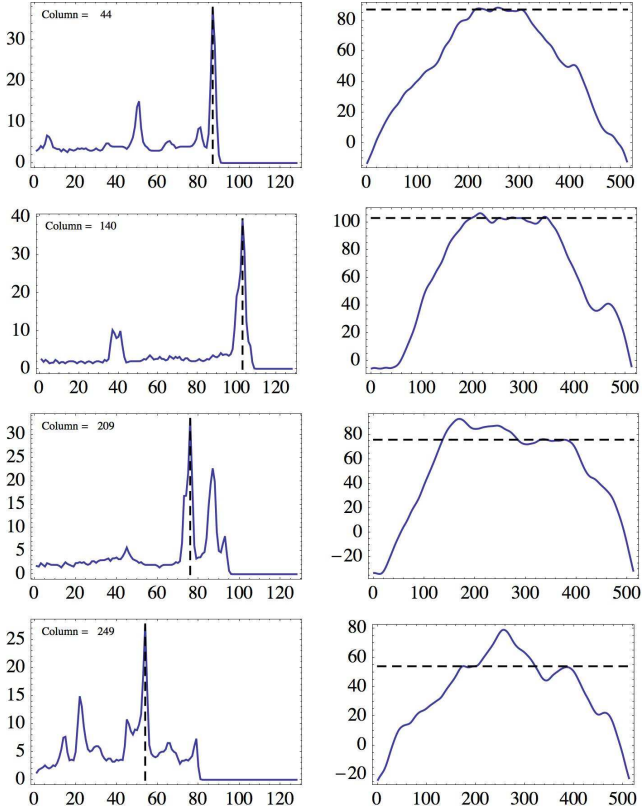}
  \caption{Profiles corresponding to the four vertical stripes shown in Fig. ~\ref{fig:LimbTM_obsmapslits}, each of them labelled by the corresponding column in the map. The left-side column shows the intensity profiles, under the assumption of a constant face-on surface brightness on the emitting sheet (the upwards direction in the map is now rightwards). The right-side column shows, instead, the profiles of the distorted limb, sectioned at the respective columns (the scales of the $x$ and $y$ coordinates are different, in order to enhance the effects of the ripples). The dashed lines refer to the position of the brightest peak in each cut.}
 \label{fig:LimbTM_gridofslits}
\end{figure}
To this purpose, let us present here some results of a simulation of the projection effect a rippled surface.
The spirit of these numerical simulations is, in a sense, like that of the model of distorted sheet presented by \citet{Hester1987}, in order to justify the observed structures present in middle-aged SNRs, and in particular in the Cygnus Loop.

While a forthcoming paper will be devoted to a more detailed statistical analysis, let us present here, for the sake of illustration, just the results of a single realization of small radial fluctuations of an otherwise spherical surface (representing the blast wave).
The radial fluctuations have been simulated with an isotropic Kolmogorov spectrum ($\dl r(k)\propto k^{-5/3}$); a separation of one decade between the lowest and the highest wavelength has been chosen to give a better by-eye match to real cases.
These radial fluctuations have been then mapped onto a portion of a sphere, and observed edge-on: the resulting projected image shows the presence of several filaments, of all intensities (see Fig. ~\ref{fig:LimbTM_obsmapslits}).
Here the local surface brightnesses are simply proportional to the total length of the layers crossing a given ray path: this is equivalent to the assumption of a constant face-on surface on this emitting sheet.

Then we have examined separately all the columns in the map (256, in the case shown), and ranked all these slices according to the highest brightness measurable in each of them.
This process has been conducted automatically, to avoid subjective choices; only at a later stage, since the brightest slices have shown the tendency to cluster spatially, we have manually selected just one slice as representative for each cluster.
The selected slices are marked by vertical segments in Fig.~\ref{fig:LimbTM_obsmapslits}, while the various panels in Fig.~\ref{fig:LimbTM_gridofslits} show the shapes of the corrugated limb in each slice: from them, it is apparent that the multiple intersections, in correspondence of the brightest projected filaments, are not the exception but rather the rule.

As a final comment, it is worth noticing that when the shock is not spatially resolved no bias is introduced in the estimation of the upstream neutral density. This is due to the fact that in the case of a corrugated shock the map shows both brighter regions and regions of depleted emission. Hence when the shock is not resolved, the total luminosity remains the same with respect to a non corrugated shock.

\section{Re-analysis of Nikolic et al. data}

The model parameters derived from the analysis of the HST image, performed in the previous section, should be tested on the data from N+13 (their Table S1) that, in spite of having a lower resolution, allow one to follow the $I_b/I_n$ spatial changes across the shock.
While the data in Table S1 are reported to be correct, the authors acknowledged in the arXiv version of the paper (arXiv:1302.4328v2) that the reference direction of the inner shock rim in Fig.~3 in N+13 was chosen incorrectly which mostly affected the Ib/In ratio trend, as plotted in that figure.
For this reason our present analysis directly refers to their Table.
In addition, the authors provided us with the measured intensities of the narrow ($I_n$) and broad ($I_b$) components separately, together with the broad-line centroid offset with respect to the narrow-line centroid.
In order to derive suitable $I_b+I_n$ and $I_b/I_n$ profiles we have proceeded as follows: first, the coordinates of the bin centroids have been rotated (by 38 degrees, anticlockwise), in order to orientate the axes respectively parallel to the filament and orthogonal to it.

Since we have noticed some trend in the brightness along the filament, in order to get a neater radial profile 
we have first corrected for a smooth trend along the filament, by fitting it with a (third degree) polynomial: this correction has been then applied just to improve the average profile for the total line intensity across the filament; while obviously the intensity ratio $I_b/I_n$ {is not affected.}

\begin{figure}
 \includegraphics[width=1.0\columnwidth,bb= 63 65 300 277, clip=true]{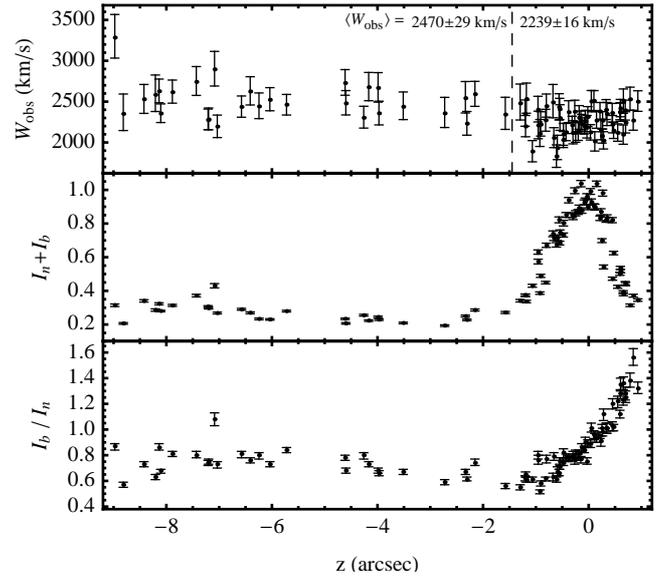}
  \caption{Profiles of the total intensity (upper panel) and of $I_b/I_n$ (lower panel), as derived from the data of N+13 (see text). The intensity profile is normalized to its peak, while the origin of the coordinate corresponds to the intensity peak.}  
 \label{fig:NikolicProfiles}
\end{figure}

Finally, in order to provide a higher homogeneity between the points used to draw the profiles, we have selected the central 15~arcsec along the filament (89 selected positions, out of the original 133).
The final results of this procedure are shown in Fig.~\ref{fig:NikolicProfiles} for $W\rs{obs}$, $I_n+I_b$, and $I_b/I_n$.
\begin{figure}
 \includegraphics[width=1.0\columnwidth,bb= 83 40 286 282, clip=true]{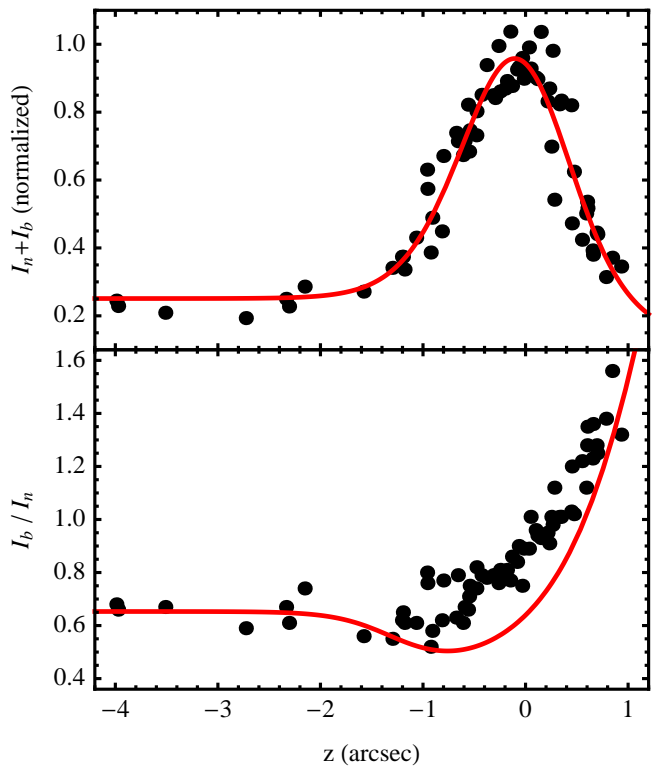}
  \caption{Simulated $I_n+I_b$ and $I_b/I_n$ profiles for the spatial resolution of N+13 (solid red lines), compared with actual observations in that work (black dots). The upper panel shows the result of the fit to the total line flux, needed to set the spatial resolution. The lower panel compares instead the prediction for the $I_b/I_n$ profile, compared with the data.}
 \label{fig:Translate_to_Nikolic}
\end{figure}

\subsection{Matching the observed $I_b/I_n$ ratio}

Let us now take the model of spatial profile of the total emission, as derived in the previous section, and downgrade it to fit the same profile, at the lower resolution as in N+13.
The best match is reached after convolving the original profile with a Gaussian function with a FWHM of 0.95~arcsec, compatible with the average spatial resolution of the N+13 data (See upper panel of Fig.~\ref{fig:Translate_to_Nikolic}; we had also to optimize spatial offset, flux scale and background level).

Then we have calculated the expected profile of $I_b/I_n$, shown in Fig.~\ref{fig:Translate_to_Nikolic}.
The main spectroscopic trend, namely the increase of $I_b/I_n$ in the projected downstream, is rather well reproduced: the asymptotic increase is well matched, and also the small dip of $I_b/I_n$ with respect to the upstream limit is marginally detected; moreover, the bending in the downstream trend as well as the start of an increase of $I_b/I_n$ before the peak in $I_b+I_n$ are clearly effects resulting from the limited spatial resolution.

The only observational effect that is not correctly reproduced by our model is the prompt increase of $I_b/I_n$ right after its minimum (at $z\simeq 1$ arcsec, in the figure).
This effect is probably the result of the superposition of layers with different densities, and therefore different $h_R$ values: while for the denser layer $I_n$ goes down promptly in the downstream, $I_b$ survives a bit longer, with the effect of slightly increasing the line components ratio, in the intermediate region.
Unfortunately, this kind of modelling is not quantitatively viable, due to the limitations of the presently available data.

\begin{figure}
\includegraphics[width=1.0\columnwidth, clip=true]{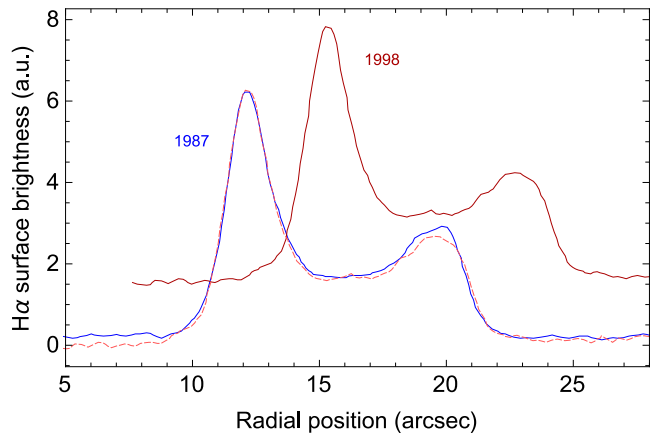}
\caption{Radial profile of the total $\Hal$ emission of a portion of northwestern filament of SN 1006 at epochs 1987.32 and 1998.48 from \protect\cite{Winkler+2003}.
The dashed line corresponds to the 1998 profile shifted backwards by 3.16 arcsec.
The shapes of the two profiles are virtually identical with the exception of a small displacement of the outer edge in the 1998 profile of $\sim0.1$ arcsec which is, however, compatible with the spatial resolution.}
\label{fig:prop_mot}
\end{figure}

\subsection{Width of the broad-line component}

Another important diagnostic tool is the width of the broad-line component, which is directly linked to the shock velocity, even though with some dependence on the thermal equilibration level \cite[see e.g. Fig.~10 in][]{paperII}.
However, N+13 showed that there is a variation of the broad line FWHM along the Balmer filament that increases from $W_b = 2357.7\U{km\,s^{-1}}$ in the inner part, up to $W_b =2555.4\U{km\,s^{-1}}$ in the outer one  (see Fig.~1 in that paper).
The average line widths given in the upper panel of Fig.~\ref{fig:NikolicProfiles}, $2239\pm16$ and $2470\pm29\U{km\,s^{-1}}$ respectively, are slightly different only because we used a different selection.
If such a variation were simply due to a different shock velocity in different parts of the shock, the required velocity difference should be $\Delta\Vsh\simeq 500\U{km\,s^{-1}}$.
But, as also noticed by N+13, this would be incompatible with the almost unchanged profile of the shock over two decades of observations. 

In order to put an upper limit on $\Delta\Vsh$ we reused the data in \cite{Winkler+2003}, taken from the sector F of their Fig.~2, which almost completely overlaps with the portion of filament analyzed by N+13.
For this sector \cite{Winkler+2003}, estimated a mean proper motion of $283.2\pm1.0\U{mas\,yr^{-1}}$, which is equivalent to a shock velocity $2685\pm9\,\dtwo\U{km\,s^{-s}}$.
We have compared the filament profiles at years 1987.32 and 1998.48, as shown in their Fig.~3), after shifting the 1998 profile backwards by 3.16~arcsec in order to match the inner peak of the 1987 profile.
As shown in Fig.~\ref{fig:prop_mot} the whole profile between the two epochs is almost unchanged.
A small displacement of $\sim0.1$~arcsec between the outer edges of the two profiles is visible which is, however, smaller than the image resolution and compatible with the pixel size of 0.1~arcsec. As a consequence the upper limit on the differential proper motion in the plane of the sky between the inner and the outer edge, taking into account also the measurement uncertainties, is $\sim 16$ mas yr$^{-1}$, which corresponds to $\Delta V_{\rm sh,\perp}\lesssim 150\,\dtwo\U{km\,s^{-1}}$.
Such a low value cannot account by itself for the observed difference in $W_b$.
In fact, even a difference of $150\U{km\,s^{-1}}$ would imply a difference in $W_b\simeq 70$--$80\U{km\,s^{-1}}$, much less than the $\simeq230\U{km\,s^{-1}}$ difference, measured between the bright filament and the outer region (see Fig.~\ref{fig:NikolicProfiles}), then strengthening the idea that a contribution of bulk velocities to the line widths at the outer edge may be substantial.

On the other hand, as we will show in Section~\ref{sec:line_width}, the larger FWHM in the outer edge can be well explained by projection effects, while the FWHM of the inner edge is the observable to be compared with the one calculated by 1-D models.
At this point we can proceed to an estimate of SN1006 shock speed using the model in Section~\ref{sec:model}. Assuming the lowest possible value of $T_e/T_p=m_e/m_p$, we can infer the lowest value for the shock speed given by equation (10) in \cite{paperIV}, which corresponds to 2599$\U{km\,s^{-1}}$. This translates into a lowest boundary for the distance $d\gtrsim 1.9$ kpc.

As for an upper bound to $\Vsh$, and therefore to the distance, one should first set an upper limit to other sinks of energy (like, for instance, an efficient CR production), and then explore what range of values for $T_e/T_i$ is consistent with the relative behaviour of broad and narrow-line components. We have found before that the observed $I_b/I_n$ ratio is consistent with $\sim10\%$ of the proton energy to be shared with electrons, then increasing the estimate of the shock speed, and therefore of the distance, by about 5\%.
Due to all this, the shock speed that that we have assumed throughout this paper, namely $3000\U{km\,s^{-1}}$, is probably too large, by about 10\%. But, in consideration of the uncertainties involved and on the negligible effects on our main conclusions, in this work we have preferred to keep this reference value for that speed, rather than attempting a combined optimization that involves also $\Vsh$.

\begin{figure}
\includegraphics[width=1.0\columnwidth, clip=true]{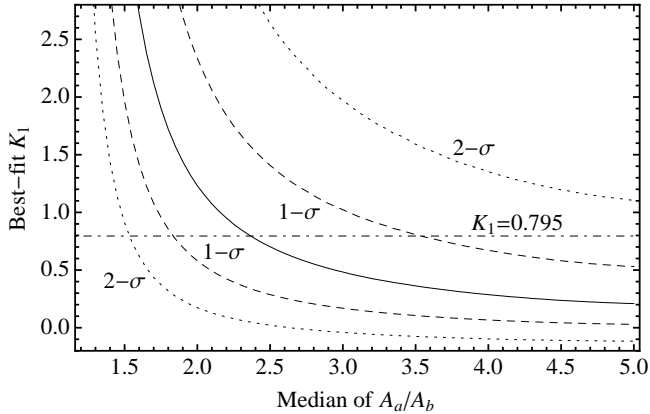}
\caption{Dependence of the best-fit value of $K_1$ on the spread of the $A_a/A_b$ distribution (expressed in terms of the median value of $A_a/A_b$, with $A_a>A_b$).}
\label{fig:K1AbovAaplane}
\end{figure}
\begin{figure}
\includegraphics[width=1.0\columnwidth, clip=true]{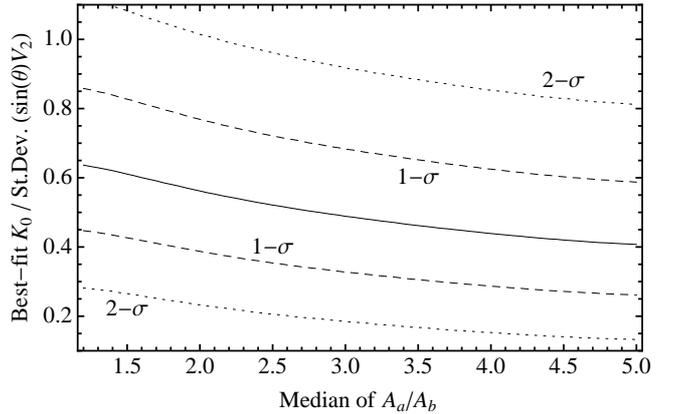}
\caption{Dependence of the best-fit value of $K_0$, scaled with the standard deviation of $\sin(\tht)\,V_2$, on the spread of the $A_a/A_b$ distribution.}
\label{fig:K0AbovAaplane}
\end{figure}

\subsection{Modelling broad-line widths and shifts} 
\label{sec:line_width}

In this section we propose a rather natural explanation for the origin of the variations in space of the width of the broad-line component ($W$), as well as for the observed shifts of the broad-line barycenter with respect to that of the narrow one ($\DlV$), both effects measured by N+13.
We show that they do not necessarily follow from changes in the physical conditions, but are more likely consequence of the geometrical structure of the shock surface.
We follow and extend a concept introduced by R+07, namely that the measured spectral width of the broad line component is the combined effect of thermal spreads and bulk motions.

Let us investigate the case in which two layers (labelled as ``$a$'' and ``$b$'') intercept the LOS.
We indicate with $A_a$ and $A_b$ their respective face-on surface brightnesses of the broad-line component, while $\tht_a$ and $\tht_b$ are their inclination angles with respect to the observer (where $\tht=0$ means a layer seen edge on).
In addition, we assume the individual layers to be thin enough to neglect their internal structure.
Finally, we assume that the velocity distribution of the hot emitting neutrals is a gaussian with an isotropic thermal dispersion $\sg_0$ and an average bulk velocity $V_2$, constant in value and always orthogonal to the shock surface.
By integrating over the radial velocity, one gets:
\begin{eqnarray}
A\rs{obs}&=&A_a/|\sin(\tht_a)|+A_b/|\sin(\tht_b)|;	\\
\label{eq:DlVobs}
\DlV\rs{obs}&=&\frac{A_a-A_b}{A\rs{obs}}V_2;	\\
\sg\rs{obs}^2&=&\sg_0^2+V_2^2\frac{A_aA_b\left(|\sin(\tht_a)|+|\sin(\tht_b)|\right)^2}{A\rs{obs}^2|\sin(\tht_a)\sin(\tht_b)|},
\end{eqnarray}
where $A\rs{obs}$ is the observed surface brightness, $\DlV\rs{obs}$ the observed velocity shift (with respect to the position of the narrow component), and $\sg\rs{obs}$ the observed gaussian dispersion (with the FWHM being $W\simeq2.35\,\sg$).
We also take $\sin(\tht_a)$ to be positive and $\sin(\tht_b)$ to be negative; namely one layer moving towards the observer and the other moving away.

It should be noticed from Eq.~\ref{eq:DlVobs} that, if $A_a=A_b$, necessarily the broad-to-narrow component velocity shift must be zero: therefore, the measurement by N+13 of both positive and negative shifts requires some spatial variability in the face-on surface brightness.

If $\tht_a$ and $\tht_b$ are close to each other, in absolute value (let us use the symbol $\tht$ for both), then we have:
\begin{eqnarray}
\label{eq:DlVobsB}
\DlV\rs{obs}&=&S_A|\sin(\tht)|V_2;	\\
\label{eq:sgobsB}
\sg\rs{obs}^2&=&\sg_0^2+(1-S_A^2)\sin(\tht)^2V_2^2,
\end{eqnarray}
where $S_A=(A_a-A_b)/(A_a+A_b)$.
Note that the lack of a clear correlation between $W$ and $\DlV$ suggests that $|\sin(\tht)|$ and $S_A$ are statistically independent quantities.

For the following analysis we use a large number of Monte Carlo simulations, in which we assume a gaussian distribution with zero mean for $\ln(A_a/A_b)$, and an independent gaussian distribution with zero mean for $\sin(\tht)V_2$.
Every time we use Eqs.~\ref{eq:DlVobsB} and \ref{eq:sgobsB} to derive $\DlV\rs{obs}^2$ and $\sg\rs{obs}^2-\sg_0^2$, respectively.
Although the distribution of points in the $\left\{\DlV\rs{obs}^2,\sg\rs{obs}^2-\sg_0^2\right\}$ parameter plane does not shown any evident correlation, one may try derive a best-fit relation $\sg\rs{obs}^2-\sg_0^2=K_0^2+K_1\DlV\rs{obs}^2$.

From simulations one may find that $K_0$ is linearly proportional to the assumed standard deviation for the stochastic variable $\sin(\tht)V_2$, while $K_1$ is independent from it, depending only on the spread of the $A_a/A_b$ distribution. We also find that the best-fit $K_1$ has typically a positive sign, and that its uncertainty could be made very small by increasing considerably the number of points.

Unfortunately, the data on which to test this trend are rather limited in number: the number of points to the left of the dashed line in Fig.~\ref{fig:NikolicProfiles} (namely excluding the bright inner filament, where the layers on the LOS are likely more than two) is only 29. With these data one finds $\sg\rs{obs}^2=(1062.)^2+0.795(\DlV\rs{obs})^2$, where both $\sg\rs{obs}$ and $\DlV\rs{obs}$ are expressed in $\U{km\,s^{-1}}$.

By performing a large number of Monte Carlo simulations, with the same number of points, we have produced Fig.~\ref{fig:K1AbovAaplane}: it shows the dependence of the best-fit value of $K_1$ on the spread of the $A_a/A_b$ distribution (expressed in terms of the median value of $A_a/A_b$, with $A_a$ taken conventionally to be the layer with the higher face-on surface brightness). In this figure, the 1--$\sg$ and 2--$\sg$ ranges compatible with the 29-point case are also displayed. Within 1--$\sg$, the observed value $K_1=0.795$ is compatible with the median of $A_a/A_b$ being in the range [1.84,3.53]; within 2--$\sg$ the median is anyway larger than 1.53 (which means that in half of the cases the layer $a$ has a face-on surface brightness at least 53\% brighter than the layer $b$). This is a further evidence of the presence of consistent surface density fluctuations along the shock surface.

Instead, for the value of $K_0$ our simulations (see Fig.~\ref{fig:K0AbovAaplane}) show that it has a 1--$\sg$ range of about 0.3--0.8 times the standard deviation of $\sin(\tht)\,V_2$, and is nearly independent of the typical value of the $A_a/A_b$ ratio.
If we assume that the broad-line width in the region of the bright limb is not appreciably affected by bulk motion effects, we can infer $\sg_0$ from it, and therefore estimate an average value for $|\sin(\tht)|$, which would give a 1-$\sg$ range for $\tht$ of about 10--30 degrees.
These values are about one order of magnitude larger than the inclination angles from the pure edge-on case, as from Table S1 in N+13, simply because they refer to different scenarios (two layers with unbalanced face-on surface brightnesses, versus a single layer).

For this range of values, a further estimate of $f_{n,u}$ can be obtained with a photometric approach: a typical surface brightness in the region between the outer limb and the bright filament is about $1.6\E{-5}\U{ph\,cm^{-2}s^{-1}arcsec^{-2}}$, so that by assuming a density of $0.033\U{cm^{-3}}$ like in the outer limb, it comes out $f_{n,u}=0.216\,|\sin(\tht)|$, which implies typical values of 0.03--0.1 for $f_{n,u}$.
All these estimates are uncertain, nonetheless they show a consistency of the present scenario.

\begin{figure}
 \includegraphics[width=\columnwidth]{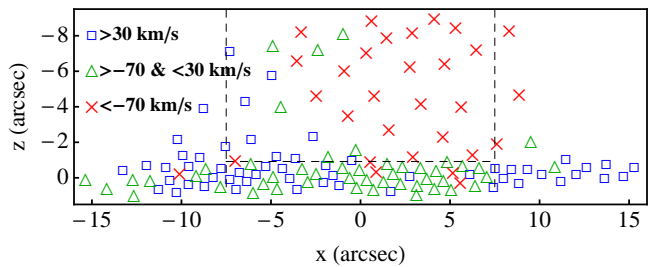}
 \caption{Map of the positions of the N+13 data points, labelled with three different symbols, depending on the measured value of the broad-line velocity offset. The figure shows some spatial coherence of the offset, with a length scale $\sim10$~arcsec. The vertical dashed lines enclose the area that we have selected for our general analysis; while the points used to investigate the relation between broad-line widths and shifts are only those above the horizontal line.}
 \label{fig:deltaVregions}
\end{figure}
Such face-on surface brightness variations should take place on rather large density scales, of at least $\sim10\arcsec$, namely $\sog0.1\U{pc}$.
A feel of this can be obtained also by looking at Fig.~\ref{fig:deltaVregions}, where the centers of the bins in which N+13 have measured $\DlV$ are displayed with three different symbols and colours.
Variations in the face-on surface brightness do not necessarily imply ambient density variations (and especially of the same level), but just variations in the density of the upstream neutral component ($n_0f_{n,u}$).

One plausible guess would be that $f_{n,u}$ is proportional to $n_0$ if the ambient gas is photoionized by the 
Galactic radiation field, so that the $n_0$ fluctuations would be  $\sim20\%$. 
However, the variations could be entirely due to $n_0$, or entirely due to $f_{n,u}$. 
In fact, if $n_0$ and $f_{n,u}$ are anti-correlated due to photoionization from the local shock itself, the density 
fluctuations could be larger.

The scenario presented in this section is in a sense similar to that discussed by \citet{Shimoda+2015}, but with opposite results.
Namely we find that, in the case of two (or more) oblique layers along the LOS, the effective width of the broad component is increasing because of the composition of the thermal width with the bulk velocities of the two components; in doing this, we have assumed that the shock velocity is always aligned with the shock normal (even though this condition does not seem to be strictly required).
Instead, in \citet{Shimoda+2015} the dominant effect is the decrease of the effective shock velocity, with respect to that estimated from astrometric measurements, because in the presence of ripples the shock is oblique almost everywhere.
A close comparison of these two results is difficult, because the conclusion of \citet{Shimoda+2015} is based on the output of a numerical simulation.
However, we believe that our approach is better justified here, because the ripples are at a rather low level (as derived from our model, as well as from a direct inspection of the edge shape).
And, in fact, the increased width outside the bright filament is more naturally explained in our scenario.
We do not exclude that, in cases in which turbulence is more important, the scheme proposed by \citet{Shimoda+2015} could be more appropriate.

\section{Comparing density estimates}

In this section we will focus on one of the main outcomes of this work, namely the estimates of the ambient density in the northwestern limb. In the above sections we have derived ambient densities ranging from an average value of $0.03\U{cm^{-3}}$ in the outer edge to about $0.10\U{cm^{-3}}$ in the bright filament. These values are a factor 3--10 lower than previous estimates that can be found in literature, hence it is worth discussing possible reasons for such a discrepancy.
A commonly accepted scenario is that SN~1006 expands in a tenuous ambient medium, with a density gradient directed towards the northwestern side, and/or with a denser cloud located near the northwestern limb. Nevertheless there is not a general agreement on the absolute value of the density.

Analyses of the X-ray thermal emission on the southeastern side lead to very low ambient densities, such as \citet{Winkler+2014} who measured $n_0=0.045_{-0.020}^{+0.049}\U{cm^{-3}}$; similar or even lower densities are found by \citet{Miceli+2012}.
With such low ambient densities it is not at all a surprise that SN~1006 is still in the ejecta-dominated (ED) regime in that region. This means that in spite of the large SNR size the total mass of the swept up material is still moderate.
Using a model from \citet{Chevalier1982}, in which the ejecta density distribution has a flat inner core and a power-law outer profile proportional to $r^{-n}$ with an index $n=7$, and assuming that the SNR is near the end of its ED phase, one can set the following upper limit for the ambient density:
\begin{equation}
n_0\leq0.035\,\dtwo^{-3}\Big(M\rs{ej}/1.4 M_\odot\Big)\U{cm^{-3}}.
\end{equation}
The Balmer filament  in the southeast region is very well approximated by a circle. With respect to this circle, the northwest filament has a radial distances smaller by about 8\%.
One could naturally explain such a distortion level by assuming that the ambient density is higher on the northwestern side, and possibly also that the shock on that side is entering the Sedov-Taylor phase. Indeed, the lower expansion velocity on that side would be consistent with a later phase \citep[see, e.g. ][]{Winkler+2014}. 
Since at a given time during the ED phase the SNR radius is proportional to $n_0^{1/n}$, for n = 7 a density factor $\simeq 2$ would be sufficient to account for the small distortion on the northwestern side. The required density difference would be even smaller if the remnant were in a later evolutionary phase.

In the case of an isolated higher density cloud, the limb distortion does not allow any stringent limit, but the shock velocity does.
Let us use the simple argument that the ram pressure in the shock downstream is everywhere a constant fraction of that upstream, so that the $V_{\rm sh} \propto n^{-1/2}$ scaling can be applied. In this case a lower expansion velocity by $\simeq40\%$ on the northwestern side would allow an overdensity no higher than a factor $\sim3$.
Since both the indentation and the lower velocity must be fitted simultaneously, a likely possibility is that both effects occur at the same time. Alternatively, a cloud with 3 times higher density would naturally account for both the indentation and the lower proper motion, provided that the blast wave reached it when the SNR was about $20\%$ smaller than its present size.
Therefore, from these kinds of arguments one should be surprised to measure ambient densities larger than $\simeq0.1\U{cm^{-3}}$ on the northwestern side.

Let us now come to direct measurements, and begin with Balmer observations: R+07, on the basis of the modelling that we have mentioned in Section 6.2, have derived $0.25\U{cm^{-3}}\leq n_0\leq0.4\U{cm^{-3}}$; while \citet{Heng+2007}, re-analyzing the same HST image, corrected the R+07 density estimate to the range 0.15--$0.3\U{cm^{-3}}$. However, we have already discussed above the reasons why this kind of analysis can be misleading.

Let us consider the results of some X-ray analyses. \citet{WinklerLong1997} derived a preshock density $\sim1.0\U{cm^{-3}}$ based on the comparison of the relative position between optical and ROSAT X-ray emission. Analyzing Chandra data \citet{Long+2003} derived an ionization time scale of $220\U{cm^{-3}\,yr}$ for the northwestern limb; measuring a thickness of $17\U{arcsec}$ for the X-ray emitting layer and assuming $(1/4)\times3000\U{km\,s^{-1}}$ for the postshock flow velocity, they estimated a characteristic time of $240\U{yr}$, and therefore an ambient density of $\sim0.25\U{cm^{-3}}$. One should notice though that, if the contact discontinuity is not too far from the forward shock the downstream velocity decreases in the downstream, so that the characteristic time must be longer and therefore the ambient density should be lower.

A similar analysis has been performed by \citet{Acero+2007}, on the basis of XMM-Newton data. An interesting aspect of that work is that it measured not only a region of the bright limb (labelled as NW-1) but also an area of faint emission in front of the bright filament (labelled as NWf). While for the bright filament an ambient density of about $0.15\U{cm^{-3}}$ was derived, the ambient density for the outer faint region was estimated to about $0.05\U{cm^{-3}}$. The authors suggested that a density of $0.05\U{cm^{-3}}$ is representative of the ambient medium around SN~1006, except for the bright northwestern filament. Our proposed scenario of a shock moving through a patchy ambient medium would be consistent with these observations\footnote{Notice that, if the ram pressure scaling law is applied to the NW filament, the upper limit on the proper motion difference between the front and back edges shown in Figure~\ref{fig:prop_mot} seems in contradiction with the measured density contrast. Nevertheless, such a scaling law cannot be used straightforwardly in the case of a curved shock front: indeed, convex or concave curvatures give rise to divergent or convergent flows immediately downstream, a fact that can produce  turbulence, affecting the velocity and pressure pattern. Hence a more detailed model is required before drawing firm conclusions.}.

Using Spitzer IR data, together with a model for the destruction of grains in the post-shock gas, \citet{Winkler+2013} derived a plasma density of about $1.0\U{cm^{-3}}$, and therefore an ambient density $\sim0.25\U{cm^{-3}}$. However, the observed IR radial profiles cannot be adequately reproduced by their model: this has been interpreted as a sign that the ambient density is not uniform. But in this case it becomes unclear how reliable the density estimate itself could be.

Rather extreme estimates are made by \citet{Dubner+2002} of nearly $0.3\U{cm^{-3}}$ for the neutral component based only on H~I radio emission (this estimate refers to the average density of the cloud that the northwestern shock has just started to interact with and not necessarily to the density of the presently shocked medium), and by \citet{Laming+1996}, $\sim0.04\U{cm^{-3}}$, based on the relative intensity of UV lines.
The latter low estimate is consistent with a more recent one by \citet{Raymond+2017}, for a nearby bow-shock like structure, grounded on the measured flux in the $\mathrm{He\,II}\,\lambda 1640$ line. One should notice however that the last two results may be affected by the fact that in both UV observations the instrument aperture intercepts only part of the emission.

To conclude, the density estimates present in the literature give a rather wide range of values and, sometimes, the assumptions to derive them are not fully justified.
On the other hand our density estimates, roughly in the range 0.03--$0.1\U{cm^{-3}}$, are based on fits of the $\Hal$ radial profiles from which a reliable physical length is deduced. Such a method is quite model insensitive; our result is only mildly dependent on model parameters as shown  in Table~\ref{tab:physparams}.
In addition, our estimates would lead to an evolutionary scenario that is in reasonable agreement with the expectations.
Finally, we note that the total density that we have derived is consistent with estimates from global models of the Galactic gas distribution \cite[see e.g.][for a review]{Ferriere2001}: at a distance of about $500\U{pc}$ from the Galactic plane (which corresponds to the location of SN~1006 for a 2 kpc distance) the mean density is about $10^{-2}\U{cm^{-3}}$ (of course with an expected wide spread on local values).

\section{Discussion and conclusions}

The main goal of this work is to show the importance to disentangle geometrical effects from the shock dynamics, in order to extract reasonable values for the physical parameters. With this in mind, the results of the present work are twofold.

On one side, we have introduced a number of techniques that are particularly effective to analyse observations of Balmer filaments when the spatial resolution is good enough to resolve the physical structure of the shock transition zone.
Such techniques are especially useful if the information on the total line emission is complemented with further information on the line profile.
To this respect, we have first approached the problem in a rather general way while, in the second part of the work, we showed how to further refine these techniques applying them to actual data from a portion of the northwestern filament of SN~1006.
In this analysis, a crucial aspect is a proper treatment of the geometry of the emitting region.
The type of bending of the shock surface may affect the observed spatial profile of the total emission, the profile of the $I_b/I_n$ ratio, and even the observed width of the broad-line component; and in some cases the structure of bending can be too complex to be modelled by a constant curvature profile.
For such complications, photometry cannot be used as the primary method to estimate densities, because of the uncertainties on the path length of the emitting region along the LOS. In addition, spectral offsets of the barycenter of the broad-line component, once combined with other information, can be used to estimate the level of density inhomogeneities, and their spatial scales, in the ambient medium reached by the shock.

On the other side, we have obtained estimates of some physical parameters for the analyzed region of the limb of SN~1006, which are not at all trivial, and that in some cases are considerably different from what has been estimated or assumed in previous works.

In the previous section we have already discussed the matter of the ambient density.
Another important parameter is the upstream neutral fraction.
Our results are consistent with low upstream neutral fractions, more likely in the range 0.01--0.1.
This range could be compared with an outcome of \citet{Ghavamian+2002}, which analyzed a very deep optical spectrum of a nearby portion of the northwestern rim of SN~1006. That observation allowed those authors to detect not only the $\Hal$ line, but also the $\Hbt$ and $\Hgm$ lines, and for the first time the $\mathrm{He\,I}\,\lambda 6678$ line and (marginally) the $\mathrm{He\,II}\,\lambda 4686$ line.
Among others, they have presented a diagnostic diagram that links the $\mathrm{He\,I}/\Hal$ ratio, the equilibration fraction, and the neutral fraction (see their Fig.~7).
On the basis of that diagram, \citet{Ghavamian+2002} estimate into 10\% the upstream neutral fraction for hydrogen, even though a neutral fraction as low as 3\% would still be compatible with the data, and than with our results.
Moreover, we notice that a low neutral fraction is indeed expected when the total density is low because, while the radiative ionization time is independent of density, the recombination time scales like the inverse of the density. In principle one could estimate the actual neutral density knowing the local radiation field, which goes, however, beyond the scope of the present work.

Finally, we have shown several pieces of evidence of the presence of consistent ambient density variations over length scales of some tenths of a pc: this follows for instance from the density estimates different by a factor $\sim3$ between the outer limb and the bright inner filament; from the measurements of offsets of the broad-line barycenter that lead to variations larger than 50\%; from the rather smooth spatial dependence of these offsets, over scales of about 0.1--0.2 pc, as well as from an observed $I_b/I_n$ spatial profile that is not well fitted by a single-density model.

If one uses the approximate proportionality $\Vsh\propto n_0^{-1/2}$ (see above) this result may seem in conflict with the stringent upper limit measured for the difference between the shock velocity in the outer limb and that in the bright filament. But this is not straightforward: being the thickness of the layer smaller than the typical length scales of the observed perturbations of the shock surface (as from the $\Hal$ image), the sound crossing time may be too long to ensure an effective pressure balance. In addition, if ambient density fluctuations (or other causes) trigger small stable oscillations about a steady state, it is consequent the instantaneous velocity differences with respect to the steady-state solution vanish when the elongation is maximum. Only a more detailed analysis, which is beyond the scopes of this work, could aim at clarifying this sort of issues.

In the present work we kept fixed the assumed shock velocity ($\Vsh$), and the level of temperature equilibration between electrons and protons ($T_e/T_p$), in order to calculate the numerical model that we have then used in our fits to the data.
We have then found a rather good match to the data so that the most likely values for $\Vsh$ and $T_e/T_p$ should not be very different from those we have used here.
A further more accurate analysis would be advisable; an analysis based on a much finer grid of numerical models, able to refine at the same time also the two quantities above.
However, in order to improve sufficiently the models one should take also into account the effect of $\Lbt$ scattering on the spatial profile of the emission of the $\Hal$ narrow line component: a full treatment of it will be presented in a forthcoming paper.
This effect, and all its consequences on spatially resolved emission models, will be the subject of a forthcoming paper.

To conclude, the physical and geometrical models discussed in the present paper are important for estimating the degree of ambiguity in the physical interpretation of existing data; but the kind of analysis presented here can develop its maximal diagnostic potential only in the presence of data with at the same time a very high spatial resolution and a sufficient spectral resolution to clearly resolve the line profile in each point.
Such data quality level will be hopefully reached in a near future using, for instance, a combination of adaptive optics and integral field spectrometry.
Another requirement for the shock transition zone to be resolved is that the SNR under investigation is a very close one  and/or expands in a low-density medium: the best such sources are SN~1006 and the Cygnus Loop, plus possibly a few other nearby Galactic SNRs.
In addition, previous works have also shown the huge diagnostic potential of combined analyses of H and He optical lines \citep[e.g.\ ][]{Ghavamian+2002}, as well as of UV lines \citep[e.g.\ ][]{Laming+1996, Raymond+2017}: an optimal diagnostic analysis should then be able to effectively combine all these different pieces of information.

\section*{Acknowledgements}
We are grateful to an anonymous referee for a careful reading of the paper and for all the suggestions that allowed us to improve the quality of the paper.
GM and RB acknowledge the financial support from the ASI-INAF agreement n. 2017-14-H.0 ``{\it Attivit\`a di Studio per la comunit\`a scientifica di Astrofisica delle Alte Energie e Fisica Astroparticellare - Analisi dati, Teoria e Simulazioni}''. SK was supported by the Ministry of Education, Science and Technological Development of the Republic of Serbia through project no. 176021, ``Visible and Invisible Matter in Nearby Galaxies: Theory and Observations''. JCR was supported by Guest Investigator grant HST-GO-13435.001 from the Space Telescope Science Institute.

Finally, we thank Anita Morlino for not being born too soon, giving us enough time to finish this work.


\end{document}